\documentclass[acmsmall]{acmart}
\AtBeginDocument{%
  }

\setcopyright{cc}
\setcctype{by}
\acmDOI{10.1145/3839487}
\acmYear{2026}
\acmJournal{PACMPL}
\acmVolume{10}
\acmNumber{OOPSLA2}
\acmArticle{355}
\acmMonth{10}
\acmSubmissionID{oopslab26main-p792-p}
\received{2026-03-17}
\received[accepted]{2026-08-06}

\usepackage{tabularx}
\usepackage{makecell}
\usepackage[dvipsnames,svgnames]{xcolor}
\usepackage{mathpartir}
\usepackage{calc}
\usepackage{wrapfig}
\usepackage{listings}
\usepackage{siunitx}

\newcommand{\code}[1]{\mbox{\texttt{#1}}}
\newcommand{\ck}{\code{cartokit}}
\newcommand{\ckhy}{\ck{}\textsubscript{\textsf{DM+NL}}}
\newcommand{\ckdm}{\ck{}}
\newcommand{\ghc}{\textsf{GitHub Copilot}}
\newcommand{\claude}{\textsf{Claude Code}}
\newcommand{\dm}{\textsf{DM}}
\newcommand{\hy}{\textsf{DM+NL}}
\newcommand{\nl}{\textsf{NL}}
\newcommand{\diff}{\code{diff}}
\newcommand{\patch}{\code{patch}}
\newcommand{\recon}{\code{recon}}

\newcommand{\blockquote}[1]{\begin{quote}
{\leftskip=-0.5cm\relax\rightskip=-0.5cm\relax
    {``}\small#1{''}
\par}
\end{quote}}

\newbool{todos}
\booltrue{todos}
\newcommand*{\todo}[2][orange]{\ifbool{todos}{{\color{#1} [#2]}}{\unskip}}

\colorlet{PromptColor}{Periwinkle}

\newcommand{\Prompt}[1]{{\color{PromptColor}\textsf{\textmd{``#1''}}}}

\definecolor{DiffColor}{HTML}{a25589}
\definecolor{cStr}{HTML}{000000}

\newcommand{\SetChannelName}{{\color{DiffColor}\textsf{setChannel}}}

\newcommand{\UnknownName}{{\color{DiffColor}\textsf{\textbf{unknown}}}}

\newcommand{\Pmapping}[1]{%
  \ifcase#1\relax
  \or PE1%   % 1
  \or PE2%   % 2
  \or PE3%   % 3
  \or PE4%   % 4
  \or PNN1%  % 5
  \or PNN2%  % 6
  \or PE5%   % 7
  \or PE6%   % 8
  \or PNN3%  % 9
  \or PNN4%  % 10
  \or PNN5%  % 11
  \or PE7%   % 12
  \or PNN6%  % 13
  \or PNN7%  % 14
  \or PE8%   % 15
  \or PNN8%  % 16
  \or PNN9%  % 17
  \or PE9%   % 18
  \fi
}

\newcommand{\p}[1]{\Pmapping{#1}}
\newcommand{\q}[1]{\textquotesingle #1\textquotesingle}

\begin{document}

\title{Direct Manipulation and Natural Language Programming, Together at Last?}

\author{Parker Ziegler}
\correspondingauthor
\orcid{0000-0001-9462-2123}
\affiliation{%
  \institution{University of California, Berkeley}
  \department{Electrical Engineering and Computer Sciences}
  \city{Berkeley}
  \country{USA}
}
\email{peziegler@cs.berkeley.edu}

\author{David Minh-Duy Cao}
\orcid{0000-0002-6163-1821}
\affiliation{%
  \institution{University of California, Berkeley}
  \department{Electrical Engineering and Computer Sciences}
  \city{Berkeley}
  \country{USA}
}
\email{dmcao@berkeley.edu}

\author{Justin Lubin}
\orcid{0000-0003-2311-1873}
\affiliation{%
  \institution{University of California, Berkeley}
  \department{Electrical Engineering and Computer Sciences}
  \city{Berkeley}
  \country{USA}
}
\email{justinlubin@berkeley.edu}

\author{Sarah E. Chasins}
\orcid{0000-0003-0557-3580}
\affiliation{%
  \institution{University of California, Berkeley}
  \department{Electrical Engineering and Computer Sciences}
  \city{Berkeley}
  \country{USA}
}
\email{schasins@cs.berkeley.edu}


\begin{abstract}
    Decades of programming languages research has contributed novel approaches to program editing that go beyond modifying text, including direct manipulation programming, structure editing, and automated refactoring tools. However, the rapid growth of natural language programming largely reinforces a view of programs as text and program editing as (unstructured) text transformation. How can we develop unified programming systems that bridge the gap between these approaches, supporting multiple editing paradigms in concert? And how would such systems change the way we program? We take a first step toward answering these questions by introducing a framework that enables program editing via both direct manipulation and natural language, and instantiate this framework in a variant of the \ck{} direct manipulation programming system (\ckhy{}). Our key insight is to treat programs as sequences of \emph{structured edits} and to use an \emph{edit language} as a shared interface for both direct manipulation and natural language interactions, leveraging constrained decoding to support the latter. Using our instantiation, we conducted a within-subjects study ($N$=18) to understand how the \emph{combination} of direct manipulation and natural language as editing modalities changes the programming process compared to each modality alone.  Perhaps surprisingly, we found that study participants overwhelmingly chose to edit via direct manipulation when both modalities were available, performing just 6.14\% of edits via natural language. Our thematic analysis of study sessions revealed that direct manipulation aided task decomposition, encouraged incremental editing, and helped mitigate known challenges in natural language programming related to understanding model capabilities and interpreting model-generated code. Conversely, natural language editing came into play largely to automate, parameterize, and replay  known edits that would otherwise be repeated tediously by hand. Our edit-based framework and study findings lay out a possible pathway for future research on programming systems that blend natural language with alternative editing modalities, building on the foundation of edit languages.
\end{abstract}




\maketitle

\section{Introduction} \label{sec:introduction}

The programming languages community has devoted decades to developing new approaches to program editing beyond text-based manipulation, from direct manipulation programming \cite{chugh_et_al_2016, hempel_chugh_2016, hempel_et_al_2019, zhang_et_al_2024} to structure editing \cite{omar_et_al_2017, adams_et_al_2025, moon_et_al_2023, teitelbaum_reps_1981} to automated code refactoring in IDEs \cite{opdyke_1992, meng_et_al_2011, thy_et_al_2023, roberts_1999}.
At the heart of these techniques is a perspective that ``programs are not [just] text'' \cite{teitelbaum_reps_1981}: they possess \emph{structure} that programming systems can (and should!) exploit to assist in the editing process.
Conversely, the recent arrival of code-generating large language models (LLMs) has largely reinforced a view of programs as text, and of program editing as a process of (unstructured) text transformation.
LLMs generate code in a purely probabilistic fashion, sampling new tokens from a distribution conditioned on the previously generated token sequence.
Recent work has attempted to reintroduce some degree of structure in this process by better contextualizing language model code completion with program information from language servers \cite{blinn_et_al_2024} and constraining language model code generation with semantic and type constraints \cite{nagy_et_al_2026, mundler_et_al_2025, wei_et_al_2023}.
Still, it remains unclear how we can integrate the potential of natural language programming and LLMs with the rich research in our community on programming paradigms that operate beyond text.

A promising approach, which we take in this paper, is to model programs not as arbitrary sequences of tokens or expressions, but as \emph{sequences of structured edits} to a starting expression $ e_0 $:
\begin{equation*}
    e = (\delta_n \circ \cdots \circ \delta_1)\, e_0
\end{equation*}
This idea is already manifest in \emph{edit languages}---such as \citet{omar_et_al_2017}'s ``actions'' in Hazel, \citet{meng_et_al_2011}'s ``edit operations'' in \textsc{Sydit}, \citet{ziegler_et_al_2025}'s ``diffs'' in \ck{}, or \citet{petricek_edwards_2025}'s ``document edit types'' in Denicek---that give precise syntax and semantics to edits, defining how programs (and their outputs) evolve at a fine level of granularity.

Given a structured language of edits, the next question is how to generate them. Prior work points to many different possibilities across diverse program editing contexts. In a structure editor like Hazel \cite{omar_et_al_2017, omar_et_al_2019}, programmer interactions at the cursor dispatch edit actions that modify the ``edit state'' (i.e., the AST) directly according to Hazelnut's action semantics. In direct manipulation programming systems like Sketch-n-Sketch \cite{chugh_et_al_2016, hempel_chugh_2016}, FuseDM \cite{zhang_et_al_2024}, and \ck{} \cite{ziegler_et_al_2025}, GUI interactions (e.g., button presses, slider drags) generate edits, which are subsequently consumed by program modification techniques (e.g., fusion \cite{zhang_et_al_2024}, \patch{}-\recon{} \cite{ziegler_et_al_2025}) that translate these edits into transformations on the program and its output. In this paper, we explore how we can extend the notion of edit sequence generation to natural language programming, using an LLM to produce \emph{structured} program edits (e.g., ``actions,'' ``edit operations,'' ``diffs'')---as opposed to \emph{unstructured} program text---from programmers' prompts.

\paragraph{Bringing Together Direct Manipulation and Natural Language Programming}

Unifying multiple edit modalities around an edit language as a shared interface opens up new directions for programming systems that support interchangeable editing across paradigms. Such programming systems, in turn, let us explore open-ended research questions about \emph{how the programming process changes} when distinct paradigms are brought together.

In this paper, we take a first step toward this vision by specifically considering two programming paradigms, direct manipulation programming and natural language programming, asking: \emph{How does the programming process change when both direct manipulation and natural language are available as editing modalities in a single programming system}? To explore this question, we first extend the \ck{} direct manipulation programming system for geospatial visualization \cite{ziegler_et_al_2025} with support for natural language editing in a variant we call \ckhy{}, using \ck{}'s existing structured \diff{}s as a shared edit language. In our variant, programmers' natural language prompts are translated by an LLM to \diff{} sequences, which can be directly interpreted by \ck{}'s existing \diff{} semantics for GUI interactions.
For example, a prompt like \Prompt{Visualize the Democratic vote share by precinct for the `Election 2024' layer, using a political color scheme} might produce the following \diff{} sequence, which will modify the marks for a particular map layer and select a data column to control the color of layer features:
{\small
\[
\begin{array}{@{}r@{\;}l@{}}
  \SetChannelName(\textcolor{cStr}{\code{\q{election\_\_2024}}}, & \textcolor{cStr}{\code{\q{fill-color-scheme}}},\; \code{(d)} \rightarrow \textcolor{cStr}{\code{\q{schemeRdBu}}} ) \\
  {}\circ\;\SetChannelName(\textcolor{cStr}{\code{\q{election\_\_2024}}}, & \textcolor{cStr}{\code{\q{fill-classification-method}}},\; \code{(d)} \rightarrow \textcolor{cStr}{\code{\q{Equal Interval}}} ) \\
  {}\circ\;\SetChannelName(\textcolor{cStr}{\code{\q{election\_\_2024}}}, & \textcolor{cStr}{\code{\q{fill-attribute}}},\; \code{(d)} \rightarrow \code{d[\textcolor{cStr}{\q{pct\_dem\_lead}}]} ) \\
  {}\circ\;\SetChannelName(\textcolor{cStr}{\code{\q{election\_\_2024}}}, & \textcolor{cStr}{\code{\q{layer-type}}},\; \code{(d)} \rightarrow \textcolor{cStr}{\code{\q{Choropleth}}} )
\end{array}
\]
}
We use constrained decoding \cite{hokamp_liu_2017, geng_et_al_2023, openai_structured_outputs} to enforce that LLM-generated \diff{} sequences are both syntactically valid and satisfy additional semantic constraints (e.g., that numeric data associated with \diff{}s fall within a certain interval, or that \diff{}s only reference existing map layers and column names). LLM-generated \diff{}s integrate directly with \ck{}'s \patch{}-\recon{} architecture; from the system's perspective, they appear as if they had been generated by GUI interactions. We then conducted a within-subjects user study ($N$=18) in which participants used (i)~our new unified direct manipulation and natural language (\hy{}) programming system, (ii)~a natural language (\nl{}) programming system (\ghc{} \cite{github_copilot} with GPT-5 \cite{gpt-5}), and (iii)~a direct manipulation (\dm{}) programming system (\ck{} \cite{ziegler_et_al_2025}) to author JavaScript programs for interactive maps.

\paragraph{Key Findings}

Our study revealed new insights on how programmers move and choose between direct manipulation and natural language editing when both are available. Surprisingly, participants overwhelmingly favored direct manipulation, performing just 6.14\% of all edits through natural language prompting. We observed that direct manipulation functionality scaffolded the programming process, guiding participants from higher-level edits to lower-level refinements and shifting the main editing challenge from recall (\emph{How do I describe my intended edit?}) to recognition (\emph{What part of the GUI can help me make my edit?}). We also found that direct manipulation encouraged incremental editing that helped participants assess correctness and develop mappings between portions of the program output and the source code. When participants did use natural language, it was primarily to scale up repetitive edits or to obtain guidance when they did not know how to achieve the edit via direct manipulation. Interestingly, direct manipulation helped mitigate key challenges in natural language programming by shaping how participants phrased prompts and providing alternative views to inspect model-generated edits. 
Cumulatively, our findings lay the foundation for understanding programming systems that blend these two editing paradigms, building on the foundation of edit languages.

\paragraph{Contributions}

In summary, we make the following contributions:

\begin{itemize}
    \item Quantitative results and a thematic analysis from a within-subjects user study exploring unified direct manipulation and natural language (\hy{}) programming, with comparisons to direct manipulation (\dm{}) programming and natural language (\nl{}) programming. To our knowledge, this is the first user study of \hy{} programming as well as the first user study of \dm{} programming.
    \item To enable the above, a framework for blending direct manipulation and natural language programming, building on the idea of structured edit languages (e.g., \diff{}s).
    \item An instantiation of this framework in \ckhy{}, a programming system that supports authoring JavaScript programs for geospatial visualization through both direct manipulation and natural language. This system is, to our knowledge, the first \emph{programming system} to support both editing modalities.
\end{itemize}

\section{Background} \label{sec:background}

\paragraph{Program Edits, Patching, and Reconciliation}

\citet{ziegler_et_al_2025} introduced \patch{}-\recon{}, an architecture for synchronizing programs and outputs in a direct manipulation programming system. The framework specifically handles cases where a user has triggered an update via a direct manipulation interaction (e.g., a button press, slider drag), and the system must modify both the program and output in response to this interaction. 
In a \patch{}-\recon{} system, each direct manipulation interaction produces a \emph{structured} program edit, called a \diff{}. Then, a system developer specifies two complementary functions that operate over \diff{}s:

\begin{enumerate}
    \item A syntactic diffing operation, \patch{} ($ \textsf{Diffs}_{\mathcal{L}} \times \textsf{Prog}_{\mathcal{L}} \rightarrow \textsf{Prog}_{\mathcal{L}} $), that takes in a \diff{} $ \delta $ and a current program $ P $ for some language $ \mathcal{L} $ and \emph{incrementally} updates $ P $  to $ P' $ based on $ \delta $.
    \item A reconciliation function, \recon{} ($\textsf{Diffs}_{\mathcal{L}} \times \textsf{Val}_{\mathcal{L}} \rightarrow \textsf{Val}_{\mathcal{L}} $), that takes in a \diff{} $ \delta $ and a current program output $ V $ and \emph{incrementally} updates $ V $ to $ V' $ based on $ \delta $.
\end{enumerate}

\noindent Finally, by proving a \emph{correspondence} between \patch{} and \recon{} with respect to all \diff{}s, a system developer can leverage \patch{}-\recon{} for parallel, incremental updates of the program and output. 
Informally, proving patch-reconciliation correspondence involves demonstrating that \recon{}'s semantic update of the output effectively emulates \patch{}'s syntactic update of the program.

The primary advantages of \patch{}-\recon{} center on performance and correctness. For long-running programs, \patch{}-\recon{} produces faster program output updates than naive whole-program re-evaluation, with speedups increasing as program run time grows.
Proving patch-reconciliation correspondence also guarantees that the current program always evaluates to the current output without actually requiring whole-program re-evaluation; in other words, all edits keep the program and output in sync.
For more details, we refer readers to \citet{ziegler_et_al_2025}.

\paragraph{What Is in a \texorpdfstring{\diff{}}{diff}?}

Each \diff{} represents a structured program edit dispatched by a direct manipulation interaction. \patch{}-\recon{} is not prescriptive about \emph{what} information should be encoded in a \diff{} or how. For example, for a given application, \diff{}s could reference tree edit operations on an AST \cite{adams_et_al_2025, omar_et_al_2017}, user actions in a record-and-replay system \cite{chasins_et_al_2015}, operations in op-based CRDTs \cite{shapiro_et_al_2011}, edits produced by the standard POSIX \textsf{diff} command \cite{posix_diff}, or any other structured information describing program edits. \diff{}s are defined by the system developer and may encode any information necessary to design \patch{} and \recon{} functions as well as prove patch-reconciliation correspondence. For this work (and our implementation, \S \ref{sec:implementation}), we reuse the instantiations of \diff{}s, \patch{}, and \recon{} in \ck{} described in \S4 of \citet{ziegler_et_al_2025} ($\diff{}_{\mathcal{L}_{ck}} $, $ \patch{}_{\mathcal{L}_{ck}} $, $ \recon{}_{\mathcal{L}_{ck}} $) with modest extensions. We detail extensions specific to this work, including both new \diff{}s and cases for the proof of patch-reconciliation correspondence, in Appendix B of the supplementary materials. For insight into the structure and scale of \ck{}-generated programs, see Appendix C, which includes an example of a \ck{} \diff{} sequence and compiled JavaScript program.

\paragraph{Constrained Decoding}

Constrained decoding (CD) \cite{geng_et_al_2023, hokamp_liu_2017, nagy_et_al_2026, mundler_et_al_2025} is a technique that enforces constraints on the sequence of tokens generated by a language model. CD works by modifying a language model's sampling procedure at inference time, preventing token selection that would violate the constraints. More concretely, CD uses a \emph{completion analysis} \cite{nagy_et_al_2026} or \emph{completion engine} \cite{mundler_et_al_2025, wei_et_al_2023} to check if a selected token could lead to a valid completion of the current token sequence and, if not, backtracks to select a different token.

\section{Unifying Direct Manipulation and Natural Language via Program Edits} \label{sec:unifying-dm-nl}

\begin{figure*}
    \centering
    \includegraphics[width=\linewidth]{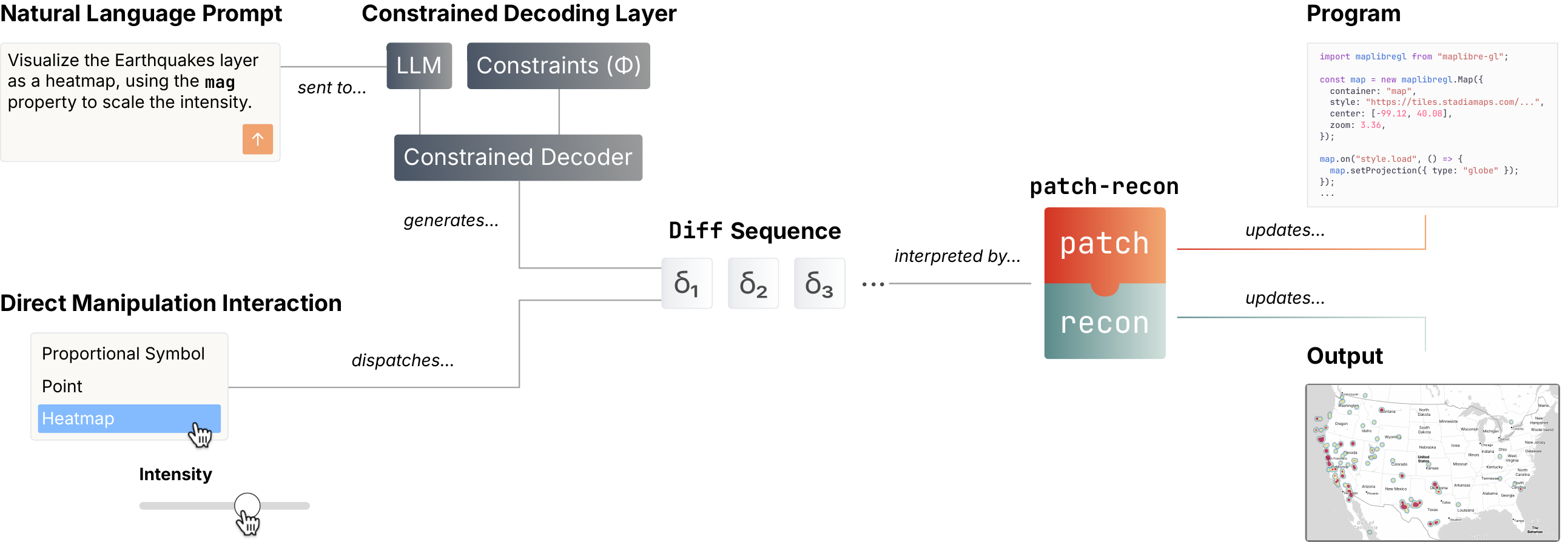}
    \caption{Generating program edits (\diff{}s) using direct manipulation
    and natural language. We use constrained decoding to translate users' natural
    language prompts, via an LLM, into structured \diff{}s that are guaranteed to
    satisfy developer-specified constraints $\Phi$. With this addition, a
    \patch{}-\recon{} system can use \emph{both} natural language and direct
    manipulation edits to produce \diff{}s; after a \diff{} is generated via either
    modality, the system applies \patch{} and \recon{} to incrementally update the
    program and output, respectively. This framework is agnostic to the design of
    \diff{}s, choice of model, choice of constraints, and choice of constrained
    decoder.}
    \Description{An overview of the constrained decoding framework. (Left) Users can apply edits either through natural language prompts (top) or direct manipulation interactions (bottom). Using constrained decoding, system developers can define syntactic and semantic constraints that define how prompts are translated to structured edit sequences (\diff{}s). (Middle) The system then applies the \patch{} and \recon{} algorithms to incrementally update the program (top right) and output (bottom right), respectively.}
    \label{fig:cd-framework}
\end{figure*}

Given a direct manipulation programming system with a definition of \diff{}s, \patch{}, and \recon{}, as well as a proof of patch-reconciliation correspondence, we can now extend the system to support program editing via natural language. The extension is straightforward: rather than using an LLM to generate new (unstructured) program text from a user's natural language prompt, we instead generate \emph{sequences of (structured) \diff{}s}. We use constrained decoding to enforce that the LLM produces only valid \diff{}s according to constraints $\Phi$ provided by the system developer. Formally:
\begin{equation*}
  cd(\mathsf{llm}, \mathsf{prompt}, \Phi) = (\delta_n \circ \cdots \circ \delta_1)
  \quad \text{such that} \quad
  \forall\, i.\; \delta_i \in \mathsf{Diffs}_{\mathcal{L}}
\end{equation*}

\noindent The key insight in this approach is to leverage \diff{}s as a \emph{shared language of program edits} that programming tools can generate from either direct manipulation edits or natural language prompts (Figure \ref{fig:cd-framework}). For a system developer, the primary additional implementation burden is to supply $\Phi$, typically as a context-free grammar or one annotated with additional semantic constraints~\cite{nagy_et_al_2026, openai_structured_outputs}. Because our framework is agnostic to the choice of constrained decoder, system developers may use the latter when syntactic constraints alone are insufficient for modeling $\textsf{Diffs}_{\mathcal{L}}$.

This approach comes with several key advantages. First, because natural language prompts produce \diff{}s, natural language edits enjoy the same performance and correctness benefits as direct manipulation edits in \patch{}-\recon{}---that is, the system can apply \diff{}s incrementally, and \diff{} application produces provably correct, parallel updates to the program and output. (Although note that model latency affects the overall performance of \diff{}s generated via natural language relative to those generated via direct manipulation, which we discuss in \S\ref{sec:switching-between-dm-and-nl}.) Second, natural language edits---again by virtue of producing \diff{}s---inherit some programming system conveniences previously only associated with direct manipulation.
For example, natural language edits can be tracked in an edit history and (incrementally) undone or redone, and their effects on interface state are immediately visible to the programmer. 
As we discuss in \S\ref{sec:experience-with-dm-affects-performance-with-nl}, this aspect of our design mitigates challenges in understanding LLM capabilities and interpreting model-generated code that occur in natural language programming settings \cite{liu_et_al_2023, nguyen_et_al_2024, zi_et_al_2025}. Finally, constraining the LLM to generate \emph{structured edits} in lieu of full programs or textual edits may offer benefits for reliability. For example, LLM-backed tools can struggle to adhere to semantic constraints on programs because they generate program text probabilistically \cite{mundler_et_al_2025, dou_et_al_2026}. By limiting generation to a fixed space of edits that enforce those semantic constraints, our framework sidesteps this problem.

\section{Implementation} \label{sec:implementation}

We extended \ck{}, a direct manipulation programming system for geospatial visualization, with support for natural language editing in a variant we call \ckhy{}. Our implementation adds only 780 lines of code (LOC) to \ck{}, which was a $\approx$13,000 LOC codebase at the time of modification. 625 LOC are TypeScript changes implementing our constrained decoding approach using Structured Outputs in the OpenAI API \cite{openai_structured_outputs} with GPT-5 \cite{gpt-5} as the model. Concretely, this involved encoding each of \ck{}'s \diff{}s as JSON Schema \cite{json_schema} using the Zod schema validation library \cite{mcdonnell_2026}. We formalized both the context-free grammar describing \ck{}'s \diff{}s and additional semantic constraints on the data associated with \diff{}s. As an example of these additional constraints, some \diff{}s in \ckhy{} support modifying \code{fill} and \code{stroke} colors of map features. To constrain GPT-5 to generating only valid 6-character hex codes for these \diff{}s, we used the following Zod encoding: \code{z.string().regex(/\textasciicircum\#[0-9A-Fa-f]\{6\}\$/)}. We provide additional details on our \diff{}-to-Zod encoding in Appendix A. The remaining 155 LOC implement Svelte \cite{svelte} components for entering prompts and displaying information from model responses.

Critically, our implementation of \ckhy{} allows full reuse of \ck{}'s existing \patch{}-\recon{} implementation; \diff{} sequences generated with GPT-5 are processed in the same manner as \diff{}s dispatched by direct manipulation interactions. This characteristic also means that natural language edits have the same guarantees around efficient incremental updates, deterministic JavaScript code generation, and the ability to be incrementally undone or redone.

\section{Experiment Design} \label{sec:experiment-design}

Having developed \ckhy{}, we set out to understand how the \emph{combination} of direct manipulation and natural language as a programming paradigm changes the programming process compared to each modality alone. Concretely, we asked the following research questions:

\begin{itemize}
    \item \textbf{RQ1} Given the task of reaching a fixed target program, what differences do we observe, if any, in participants' success rate or completion time when both direct manipulation and natural language are available as editing modalities compared to each modality alone?
    \item \textbf{RQ2} How does the combination of direct manipulation and natural language as editing modalities influence how participants edit programs, write prompts, and assess program correctness relative to each modality alone?
        \begin{itemize}
            \item[$\hookrightarrow$] \textbf{RQ2a} Do we observe differences in participants' approaches based on prior programming experience?
        \end{itemize}
\end{itemize}

\noindent We addressed these questions through a within-subjects lab study with $N$=18 geospatial data users programming interactive maps using three tools: \ckhy{} (\hy{}), \ckdm{} (\dm{} only), and \ghc{} (\nl{} only). We evaluated \textbf{RQ1} quantitatively by measuring task completion and time-on-task, and by analyzing telemetry data collected in each study condition. We assessed \textbf{RQ2} and \textbf{RQ2a} qualitatively through an inductive, reflexive thematic analysis \cite{braun_clarke_2006} of study sessions.

Although these measures are similar to those used in traditional productivity lab studies, we want to emphasize that the goal of our experiment was not to provide any definitive answer on which of the three programming paradigms is ``better'' for developer output. Rather, we sought to characterize programming behaviors in the novel multimodal editing case made possible by our system design, using the other conditions as points of comparison for contextualizing our findings.

\subsection{Participants} \label{sec:participants}

We conducted live study sessions with 18 (one transmasculine, five female, ten male, two who did not disclose) participants, the majority of whom ($66.\overline{6}\%$) were professional data journalists. We focused on data journalists as a study population for two reasons. First, many data journalists interact regularly with geospatial data and produce interactive maps as part of their work, making them target users for programming systems (like ours) that can assist in this process. Second, many data journalists come from ``non-traditional'' computing backgrounds (i.e., they may not have developed programming experience through formal computer science education). Given that direct manipulation and natural language programming systems are often designed specifically to support such users, data journalists are a relevant community for understanding how these paradigms (and their combination) affect the programming process. We supplemented our participant pool with students and researchers in journalism, Earth sciences, and geography from multiple research universities in the United States. We recruited participants through the first author's professional networks, a Slack workspace for data journalists, and university mailing lists. Table \ref{tab:participants} provides more detail on each participant's background.

To help answer \textbf{RQ2a}, we used a screening survey to intentionally recruit an equal number of expert and near-novice programmers based on self-reported experience and background with JavaScript programming and JavaScript mapping libraries. We considered participants with at least three years of JavaScript programming experience and at least two years of programming experience with JavaScript mapping libraries to be experts. Conversely, near-novices self-reported between zero and two years of programming experience with JavaScript and less than one year of programming experience with JavaScript mapping libraries. \p{10} was the only participant who straddled these classification criteria. Despite having JavaScript experience, they noted that their use of JavaScript was very sporadic. Given this fact---in concert with their zero years of experience with JavaScript mapping libraries---we chose to include them in the near-novice group.

\paragraph{Domain Expertise} 

Importantly, our near-novice categorization is not indicative of participants' experience working with geospatial data or their background in cartography. 
In fact, many of the near-novice programmers in our participant pool are expert cartographers with deep expertise in geographic information systems (GISs), geospatial analysis, and cartographic design; they simply do not currently use JavaScript to do this work.  However, the goal of this experiment is to understand direct manipulation and natural language as paradigms for \emph{programming} systems, so our expert vs. near-novice labels refer only to programming expertise.

\begin{table}
    \centering
    \caption{Backgrounds of study participants. ``NN'' in the \textbf{ID} column denotes near-novice participants (top); ``E'' denotes expert participants (bottom). \textbf{Freq. LLM Use} shows participants' self-reported frequency of LLM use for any programming activity (Never, Occasionally, Once a month, Once a week, Daily).}
    \vspace{-1em}
    \begin{tabularx}{\textwidth}{l|r|r|l|X}
        \toprule
        \thead{\footnotesize{\textbf{ID}}} & \thead{\footnotesize{\textbf{JS Exp.}} (Yrs.)} & \thead{\footnotesize{\textbf{JS Map. Exp.}} (Yrs.)} & \thead{\footnotesize{\textbf{Freq. LLM Use}}} & \thead{\footnotesize\textbf{{Occupation}}} \\
        \midrule
        \footnotesize{\p{5}} & \footnotesize{1} & \footnotesize{<1} & \footnotesize{Never} & \footnotesize{Data Journalist} \\
        \footnotesize{\p{6}} & \footnotesize{<1} & \footnotesize{<1} & \footnotesize{Occasionally} & \footnotesize{Graduate Student (Journalism)} \\
        \footnotesize{\p{9}} & \footnotesize{0} & \footnotesize{0} & \footnotesize{Occasionally} & \footnotesize{Data Journalist} \\
        \footnotesize{\p{10}} & \footnotesize{4} & \footnotesize{0} & \footnotesize{Daily} & \footnotesize{Graduate Student (Earth Sciences)} \\
        \footnotesize{\p{11}} & \footnotesize{2} & \footnotesize{0} & \footnotesize{Daily} & \footnotesize{Postdoctoral Scholar (Geography)} \\
        \footnotesize{\p{13}} & \footnotesize{2} & \footnotesize{0} & \footnotesize{Never} & \footnotesize{Data Journalist} \\
        \footnotesize{\p{14}} & \footnotesize{<1} & \footnotesize{<1} & \footnotesize{Once a week} & \footnotesize{Undergraduate Student (Comp. Sci., Geography)} \\
        \footnotesize{\p{16}} & \footnotesize{<1} & \footnotesize{0} & \footnotesize{Never} & \footnotesize{Graduate Student (Journalism)} \\
        \footnotesize{\p{17}} & \footnotesize{<1} & \footnotesize{<1} & \footnotesize{Daily} & \footnotesize{Graduate Student (Journalism)} \\
        \midrule
        \footnotesize{\p{1}} & \footnotesize{5} & \footnotesize{4} & \footnotesize{Never} & \footnotesize{Data Journalist} \\
        \footnotesize{\p{2}} & \footnotesize{10} & \footnotesize{5} & \footnotesize{Daily} & \footnotesize{Data Journalist} \\
        \footnotesize{\p{3}} & \footnotesize{14} & \footnotesize{14} & \footnotesize{Never} & \footnotesize{Data Journalist} \\
        \footnotesize{\p{4}} & \footnotesize{15} & \footnotesize{13} & \footnotesize{Once a month} & \footnotesize{Data Journalist} \\
        \footnotesize{\p{7}} & \footnotesize{3} & \footnotesize{2} & \footnotesize{Once a week} & \footnotesize{Data Journalist} \\
        \footnotesize{\p{8}} & \footnotesize{4} & \footnotesize{3} & \footnotesize{Never} & \footnotesize{Data Journalist} \\
        \footnotesize{\p{12}} & \footnotesize{15} & \footnotesize{13} & \footnotesize{Daily} & \footnotesize{Data Journalist} \\
        \footnotesize{\p{15}} & \footnotesize{7} & \footnotesize{7} & \footnotesize{Never} & \footnotesize{Data Journalist} \\
        \footnotesize{\p{18}} & \footnotesize{5} & \footnotesize{5} & \footnotesize{Occasionally} & \footnotesize{Data Journalist} \\
        \bottomrule
    \end{tabularx}
    \label{tab:participants}
    \vspace{-1em}
\end{table}

\subsection{Study Conditions and Protocol} \label{sec:study-conditions}

Our within-subjects study included three conditions:

\begin{enumerate}
    \item \textbf{\hy{} Condition}: \ckhy{}.
    \item \textbf{\dm{} Condition}: \ckdm{} \cite{ziegler_et_al_2025}.
    \item \textbf{\nl{} Condition}: \ghc{} with GPT-5. We gave participants access to \ghc{}'s web interface \cite{github_copilot} using GPT-5 \cite{gpt-5} as the underlying model.
\end{enumerate}

Within a given condition, each participant completed the following procedure.

\paragraph{Tutorial (\texorpdfstring{5\,}{5}min)} Before starting the tasks, the participant read a short tutorial highlighting key features of the interface they would be using. Tutorials were not step-by-step guides or walkthroughs of example scenarios; instead, tutorials focused on how to access various features of each interface. We provide all three tutorials in the supplementary materials.

\paragraph{Task A, \texorpdfstring{\textsc{Reproduction}}{Reproduction} (\texorpdfstring{30\,}{30}min)} In Task A, we asked participants to develop a JavaScript program to reproduce a specific map published in The Washington Post. With three conditions, each participant attempted to reproduce three different target maps---one per condition---over the course of the study. We assigned map-condition pairings in counterbalanced order to control for learning effects; see \textit{Ordering} below.
We provided participants with a static image of the target map, the URL of the news article containing the map, light instructions about the features of the map to reproduce, and the URL of the GeoJSON \cite{geojson} data to use for the task.
Before conducting the study, we tested whether using the full task description text as a prompt in the two natural language conditions was sufficient for completing the task; in our testing, it was not.
All GeoJSON datasets were stored in a public Cloudflare R2 bucket and available for download to the participants' machines. To complete the task in each condition, we required participants to execute code in StackBlitz \cite{stackblitz}, an online JavaScript IDE. We added this requirement to ensure that generated code (i)~executed without syntactic or semantic errors and (ii)~produced an output map meeting the minimum criteria for reproduction according to our rubric. Participants could abandon the task at any time.

\paragraph{Task B, \texorpdfstring{\textsc{Exploration}}{Exploration} (\texorpdfstring{10\,}{10}min)} In Task B, we asked participants to explore two geospatial datasets with the goal of authoring a JavaScript program that produced a map they considered ``desirable.'' As in Task A, we provided participants with URLs for the two GeoJSON datasets, assigned in counterbalanced order; participants could choose to use one or both datasets in their final map. Participants could end the task early by declaring they had reached a desirable map or by abandoning the task.
Note that there are no quantitative measures associated with Task B.
This task augments the qualitative portion of our data analysis, offering a
part of the session where we could observe how the programming paradigms
affected how participants engaged in exploratory programming work, without
the constraint of a researcher-assigned target output.

\paragraph{Ordering} Map-condition pairings were assigned using a counterbalanced Latin square. With three conditions, we have six possible orderings of conditions. Thus, with 18 participants, each distinct ordering was completed by a group of three participants.

\paragraph{NASA-TLX} After completing each task, participants completed the NASA Task Load Index (NASA-TLX) \cite{hart_staveland_1988} to report perceived workload for each condition-task pairing. Following guidance and findings from prior work \cite{bustamante_spain_2008, hendy_et_al_1993}, and to reduce burden on participants, we chose to remove subscale weighting that requires participants to perform many pairwise comparisons on the perceived importance of each subscale. Instead, each participant indicated perceived workload along each subscale, and we take the unweighted mean of these values to estimate overall workload. This variation is sometimes referred to as Raw TLX (RTLX) \cite{hart_2006}.

\paragraph{Post-Study Interview and Survey (\texorpdfstring{10--15\,}{10–15}min)} After completing the two tasks in each condition (2 \textsc{Tasks} $/$ \textsc{Condition} $\times$ 3 \textsc{Conditions} = 6 \textsc{Tasks}), participants reflected on their experiences in each condition in a semi-structured interview. Discussion centered on how each condition directed participants' attention while programming, how participants assessed program correctness, strategies for prompting, and strategies for navigating each interface, among other behaviors.

\paragraph{Session Duration and Compensation} Participants took part in the study remotely over Zoom over the course of 2--2.5 hours. Each participant shared their screen with the experimenter during the session, and we recorded both their screen and audio. In addition, we instrumented \ckhy{} to record anonymized \diff{} sequences triggered by direct manipulation, as well as prompts and \diff{} sequences triggered by natural language edits. Likewise, we instrumented \ckdm{} to record anonymized \diff{} sequences capturing direct manipulation interactions. Participants received compensation in the form of a \$50 gift card or a \$50 donation to a 501(c)(3) of their choice.

\subsection{Analysis} \label{sec:analysis}

\paragraph{Quantitative Analysis}

For each \textsc{Reproduction} task, we collected data on completion (\textsf{True}/\textsf{False}) and completion time. Determining completion on a reproduction task like ours can be challenging, as the experimenter's subjective assessment of the faithfulness and accuracy of a participant's map against the target map can be biased. Therefore, we used a rubric with a minimum set of unambiguous criteria that participants should implement to declare the task complete. We did not give participants access to the rubric as we suspected it could aid them in task decomposition, artificially simplifying the task. Instead, participants notified the researcher when they believed they had reproduced the map closely enough. In the event that a participant had not implemented criteria from our rubric, we asked them specifically to implement the functionality. To determine the completion time for the task, we used the timestamp from the video recordings when the participant's JavaScript program successfully executed in StackBlitz and rendered their map.

\paragraph{Qualitative Analysis}

In total, we collected 34 hours of session recordings. We then conducted an inductive, reflexive thematic analysis \cite{braun_clarke_2006} of the video data using the qualitative coding tool MaxQDA \cite{maxqda}. We started with an open coding phase in which the first two authors each independently tagged segments of the session recordings with short, descriptive phrases of participant behaviors. In a second phase, the first two authors met to group their respective open codes into a hierarchy of axial codes. Finally, the entire research team met several times to discuss, refine, and regroup these axial codes into top-level themes, discussed in \S\ref{sec:qualitative-results}.

In addition to the session recordings, we recorded and analyzed all prompts issued by participants in the \hy{} (\ckhy{}) and \nl{} (\ghc{}) conditions, as well as telemetry data of direct manipulation interactions in the \hy{} and \dm{} (\ckdm{}) conditions. We present qualitative findings on participant prompts and \diff{} sequences throughout \S\ref{sec:qualitative-results}.

\section{Quantitative Results} \label{sec:quantitative-results}

In this section, we provide quantitative results on task completion rates and times across conditions. We also present self-reported NASA Task Load Index scores for each condition-task pair.

\subsection{Task Completion and Timing} \label{sec:task-completion-and-timing}

\paragraph{\texorpdfstring{\textbf{RQ1}}{RQ1} Given the task of reaching a fixed target program, what differences do we observe, if any, in participants' success rate or completion time when both direct manipulation and natural language are available as editing modalities compared to each modality alone?}

All 18 participants (100\%) completed the \textsc{Reproduction} task in the \hy{} condition with \ckhy{}. This result was mirrored in the \dm{} condition (\ckdm{}), with all 18 participants completing (100\%). In the \nl{} condition (\ghc{}), nine out of 18 participants (50\%) completed the task, with six choosing to abandon the task before the 30-minute threshold.

\begin{figure*}[tbp]
    \centering
    \includegraphics[width=0.9\linewidth]{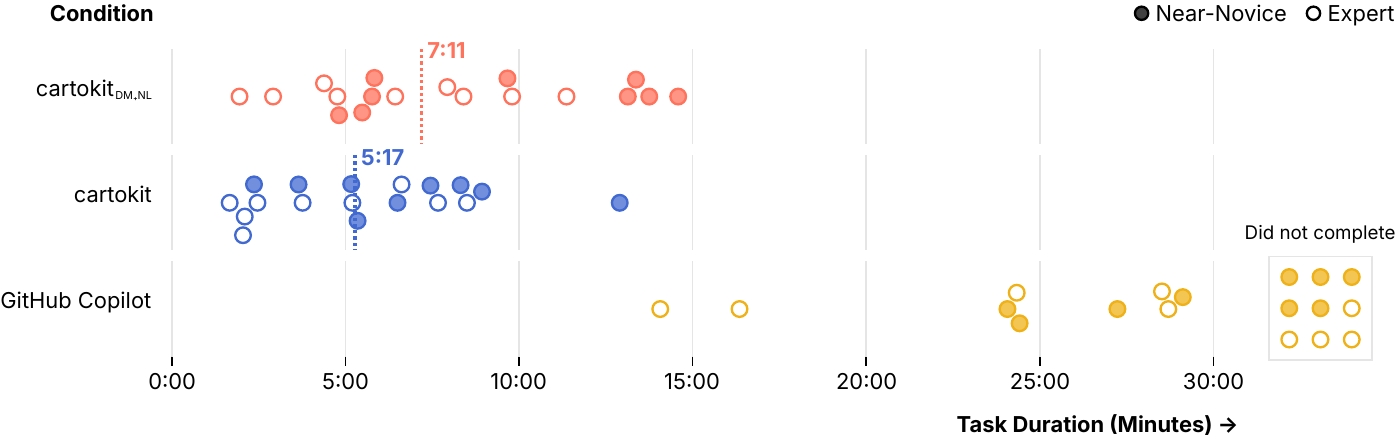}
    \caption{Completion times of participants on \textsc{Reproduction} tasks. Dashed lines and their associated labels denote median completion times across all participants for the condition. \ghc{} had a completion rate of 50\%; the nine participants who did not complete the task are reflected in the box at right.}
    \Description{A faceted beeswarm plot showing completion times of participants on \textsc{Reproduction} tasks by condition. Dashed lines show the median completion time for \ckhy{} (7m11s) and \ckdm{} (5m17s). Nine participants did not complete the \textsc{Reproduction} task; they are represented in a box at right labeled ``Did not complete''.}
    \label{fig:completion-times}
    \vspace{-0.5em}
\end{figure*}

The median completion time for the \textsc{Reproduction} task in the \hy{} condition was 7m11s. This was slightly slower than the median completion time in the \dm{} condition, 5m17s. As mentioned above, only nine of 18 participants completed the task in the \nl{} condition. Given that this data is right censored at exactly 50\% completion, we can only estimate the median completion time. To do this, we conducted a survival analysis with a Kaplan-Meier estimator \cite{kaplan_meier_1958} and used Greenwood's formula \cite{greenwood_1926} to derive a 95\% confidence interval on the median. The estimated median completion time exceeds the timing threshold (\qty{30}{min}) and is thus reported as >30m with a 95\% CI = [24m25s, >30m). Figure \ref{fig:completion-times} shows the completion times of all participants across \textsc{Reproduction} tasks.

\paragraph{Effect Size}

\begin{figure}[htbp]
  \begin{minipage}[c]{0.7\textwidth}
    \includegraphics[width=\textwidth]{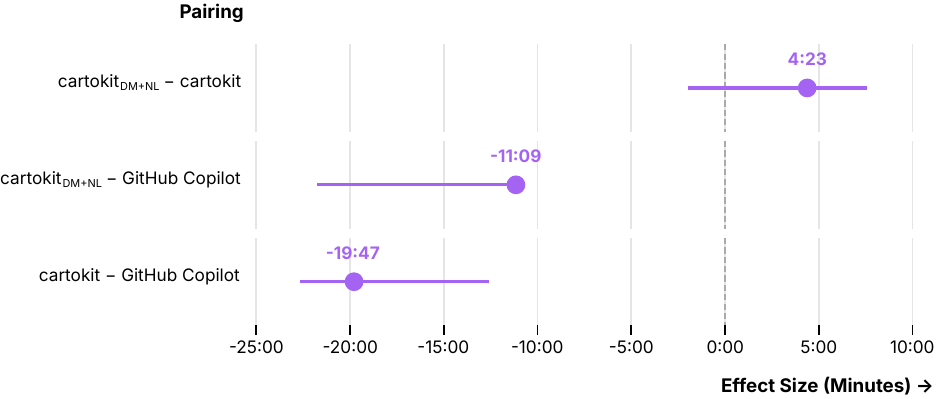}
  \end{minipage}%
  \hfill
  \begin{minipage}[c]{0.27\textwidth}
    \caption{Median difference in completion times (effect size) between conditions on \textsc{Reproduction} tasks. Negative values indicate participants were faster in the LHS condition than the RHS condition (and vice versa). Purple lines show 95\% CIs.}
    \label{fig:effect-size}
  \end{minipage}
  \Description{A forest plot showing the effect size (in minutes) between each pairing of conditions.}
  \vspace{-1em}
\end{figure}

To compute an effect size between each pair of conditions, we conducted a second survival analysis on the pairwise differences, per participant, in completion times. We then bootstrapped our Kaplan-Meier estimator using the bias corrected and accelerated bootstrap \cite{efron_1987} to derive 95\% confidence intervals. Figure \ref{fig:effect-size} shows the results. Participants were 4m23s slower (95\% CI = [-1m58s, 7m36s]) using \ckhy{} than \ckdm{}, 11m9s faster (95\% CI = [-21m45s, -10m40s]) using \ckhy{} than \ghc{}, and 19m47s faster (95\% CI = [-22m40s, -12m35s]) using \ckdm{} than \ghc{}.

\subsection{NASA Task Load Index} \label{sec:nasa-tlx}

\begin{figure*}
    \centering
    \includegraphics[width=0.925\linewidth]{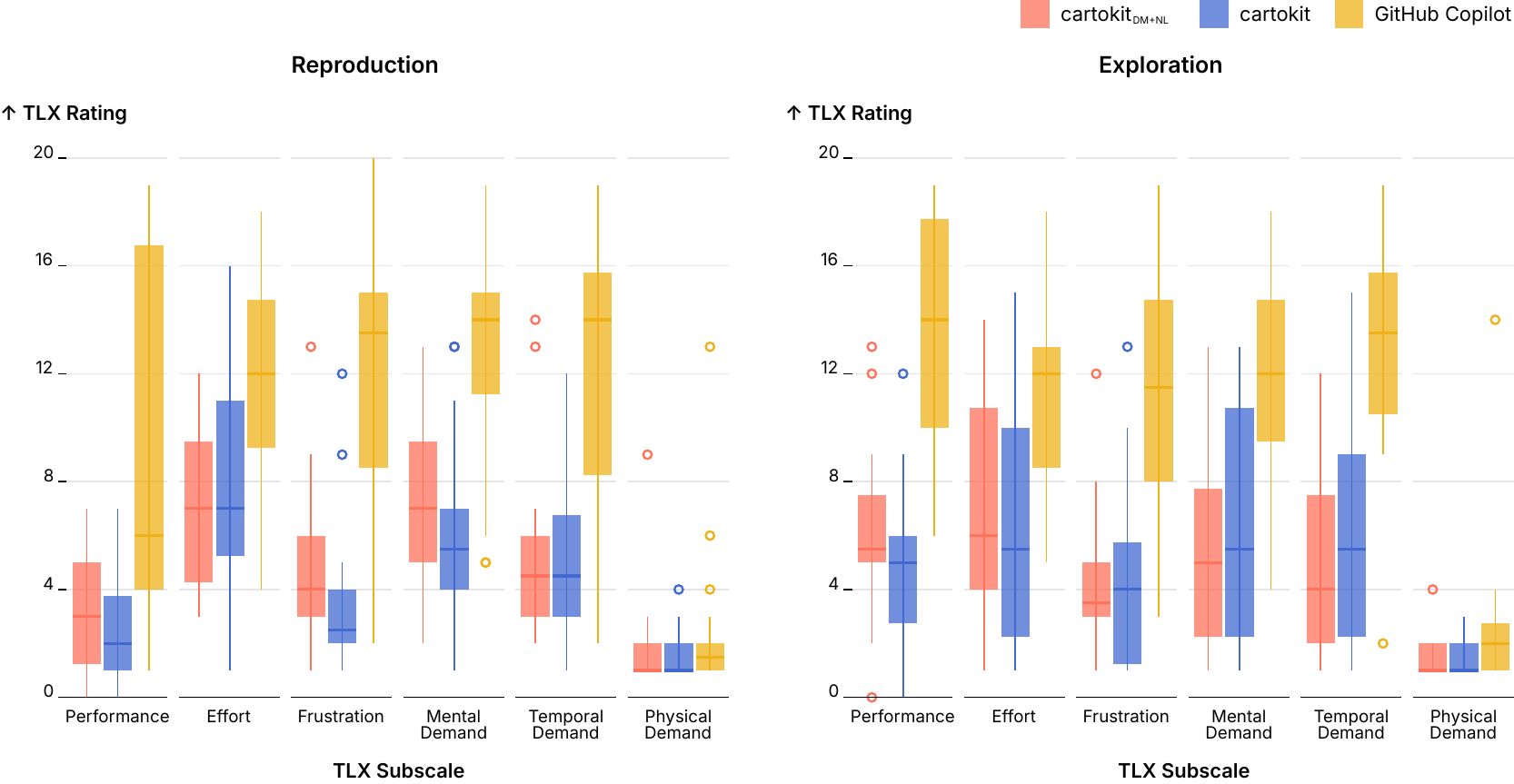}
    \caption{Distribution of participant NASA-TLX scores by subscale and condition, on \textsc{Reproduction} tasks (left) and \textsc{Exploration} tasks (right). Lower values indicate less perceived workload or, in the case of the Performance subscale, better performance.}
    \Description{Box plots showing distributions of participant NASA-TLX scores by subscale and condition, on \textsc{Reproduction} tasks (left) and \textsc{Exploration} tasks (right).}
    \label{fig:tlx}
    \vspace{-1em}
\end{figure*}

Participants self-reported their perceived workload after each \textsc{Reproduction} and \textsc{Exploration} task, using ratings from 0 to 20 along the six subscales used in the NASA-TLX: Performance, Effort, Frustration, Mental Demand, Temporal Demand, and Physical Demand. Figure \ref{fig:tlx} shows the distributions of participant ratings by subscale. Lower values denote less perceived workload or, in the case of the Performance subscale, \emph{better} perceived performance. We then computed the unweighted mean of these values as the composite Raw TLX score.

On \textsc{Reproduction} tasks, participants self-reported a mean Raw TLX score of 4.86 (SE $\pm$0.40) with \ckhy{}, 4.33$\pm$0.51 with \ckdm{}, and 10.02$\pm$0.71 with \ghc{}. On \textsc{Exploration} tasks, participants self-reported a mean Raw TLX score of 4.95$\pm$0.58 with \ckhy{}, 4.88$\pm$0.67 with \ckdm{}, and 10.51$\pm$0.63 with \ghc{}. In addition, we compare the differences in mean Raw TLX scores between conditions per task type, reporting 95\% confidence intervals using the bias corrected and accelerated bootstrap (Table \ref{tab:tlx}). Our results show that participants' Raw TLX scores in the \hy{} condition were not statistically different from those in the \dm{} condition (regardless of task type), suggesting that the addition of natural language editing capabilities in \ckhy{} was not meaningfully associated with an increase or reduction in participants' perceived workload. However, Raw TLX scores in the \hy{} condition were markedly lower than those in the \nl{} condition, suggesting that the presence of direct manipulation editing capabilities may have helped alleviate some perceived workload associated with natural language-only editing.

\begin{table}[htbp]
    \centering
    \caption{Differences in means of participant Raw TLX scores, paired by condition and task type. Negative values indicate participants' perceived workload on the LHS condition was lower than on the RHS condition.}
    \vspace{-1em}
    \begin{tabularx}{\textwidth}{X|X|X}
        \toprule
        \thead{\footnotesize{\textbf{Pairing}}} & \thead{\footnotesize{\textbf{Diff. Mean Raw TLX [95\% CI]}} \\ \footnotesize{\textsc{Reproduction}}} & \thead{\footnotesize{\textbf{Diff. Mean Raw TLX [95\% CI]}} \\ \footnotesize{\textsc{Exploration}}} \\
        \midrule
        \footnotesize{\ckhy{} — \ckdm{}} & \footnotesize{0.53 [-0.94, 1.44]} & \footnotesize{0.07 [-0.72, 0.95]} \\
        \footnotesize{\ckhy{} — \ghc{}} & \footnotesize{-5.16 [-6.44, -3.53]} & \footnotesize{-5.56 [-7.17, -4.06]} \\
        \footnotesize{\ckdm{} — \ghc{}} & \footnotesize{-5.69 [-7.08, -3.91]} & \footnotesize{-5.63 [-7.28, -4.20]} \\
        \bottomrule
    \end{tabularx}
    \label{tab:tlx}
    \vspace{-0.75em}
\end{table}

\section{Qualitative Results} \label{sec:qualitative-results}

In this section, we discuss qualitative findings from our within-subjects study, focusing in particular on how the \hy{} condition shaped participants' programming process relative to the \dm{} and \nl{} conditions (\textbf{RQ2}), and how behaviors differed between near-novices and experts (\textbf{RQ2a}). Throughout this section, we make references to participants' prompts, denoted by {\color{PromptColor}\textsf{periwinkle}} text.

\subsection{How Did Participants Edit Programs?}

\subsubsection{In the \texorpdfstring{\hy{}}{DM+NL} Condition, Participants Overwhelmingly Used Direct Manipulation Editing, but Experimented with Natural Language Editing to Automate Repetitive or Long-Range Tasks} \label{sec:preferring-dm}

In total, participants issued 55 prompts across all tasks in the \hy{} condition, amounting to $\approx$1.5 prompts per participant per task. Of the 2,946 \diff{}s dispatched by participants in this condition, only 181 ($\approx$6.14\%) were natural language edits. Qualitative analysis of participant behaviors largely reflected this skew. We observed that the majority (14/18) of participants started their editing process with direct manipulation, with only four participants opting to start with prompting; some participants never prompted in the course of a task. When asked about avoiding natural language editing, participants explained that they often already understood how to achieve their desired edit through direct manipulation, so it ``didn't feel faster'' (\p{1})---and in fact, it ``felt like I'm adding layers of abstraction to what I'm trying to do'' (\p{4})---to describe the edit in natural language. Among participants who did prompt, we observed that they primarily leveraged natural language editing to (i)~automate repetitive tasks that would be tedious to perform via direct manipulation or (ii)~scaffold long-range edit sequences. \S{\ref{sec:switching-between-dm-and-nl}} dives into these behaviors, providing a detailed analysis of when, how, and why participants switched between direct manipulation and natural language.

\subsubsection{In the \texorpdfstring{\hy{}}{DM+NL} and \texorpdfstring{\dm{}}{DM} Conditions, the GUI Scaffolded How Participants Edited Programs, Guiding Them from \texorpdfstring{``High-Abstraction''}{"High-Abstraction"} to \texorpdfstring{``Low-Abstraction''}{"Low-Abstraction"} Edits} \label{sec:scaffolding-edits}

Participants were remarkably consistent in their approach to editing programs in both \ckhy{} and \ckdm{}, with the majority turning first to GUI controls supporting high-abstraction edits, such as modifying layer types and encoding channels, and moving progressively toward controls supporting low-abstraction edits, like adjusting colors, opacities, and symbol size ranges. The former category of edits is ``high-abstraction'' in the sense that they abstract over many concrete details of the data encoding strategy, whereas the latter correspond to styling ``tweaks'' of specific visual properties. While the GUI made performing both classes of edits similarly efficient for participants, their impacts on the program were quite different. High-abstraction edits often introduced larger, non-local changes to the program, such as introducing new function definitions and call sites to transform or analyze data, whereas low-abstraction edits had localized effects on individual strings or number literals.

We hypothesize that this high-abstraction to low-abstraction editing pattern reflects how the GUI in \ckhy{} and \ckdm{} both decomposed and scaffolded the programming process for participants. On \textsc{Reproduction} tasks, many participants (14) navigated the layer editing GUI sequentially, treating each GUI control as a distinct editing choice as they moved closer to the target map. On \textsc{Exploration} tasks, participants made heavy use of the top-level \textsf{Layer Type} controls to transition quickly between standard geospatial visualizations, then shifted to iteratively tweaking details like color schemes or symbol sizing as they refined their maps. Importantly, the GUI itself helped to communicate the kinds of edits that were possible, shifting participants' main programming task from recall (of syntax) to recognition (of GUI controls) \cite{budiu_2026}. In addition, by implicitly guiding participants through a sequence of editing decisions, the GUI absorbed much of the problem decomposition burden that prior work has shown to be a central obstacle in natural language programming, particularly for near-novices \cite{kazemitabaar_et_al_2024, babe_et_al_2024, nguyen_et_al_2024}. We suspect that this was a key factor in participants' strong performance in both the \hy{} and \dm{} conditions.

\paragraph{\texorpdfstring{\textbf{Comparison}}{Comparison}: In the \texorpdfstring{$\nl{}$}{NL} condition, participant prompts attempted to perform large-scale changes all at once, which could lead to erring programs or regressions in functionality}

Participants regularly issued prompts that tried to make entire maps in ``one shot.'' Even in instances where they effectively decomposed a target map into discrete components (e.g., layers, projections, data encodings), going ``too big'' (\p{15}) in any single prompt could result in updated programs that regressed in functionality (\p{7}), removed desired functionality (\p{4}, \p{12}), or introduced runtime errors that affected the program output (\p{13}, \p{16}, \p{8}, \p{15}). For example, \p{8} was making steady progress toward completing a \textsc{Reproduction} task, with both provided datasets successfully rendering on the map. They then moved to specify the size and color encodings of features in one of the layers (\Prompt{Scale each square by abundance and color by population change.}). This prompt resulted in a JavaScript program that triggered an out-of-memory exception at run time. Concerned, \p{8} issued a follow-up prompt to complete a simpler part of the task (\Prompt{color the range as grey}), explaining, ``I want to get back to a working map after I messed it up, then I'll work on the color scheme.'' However, this prompt resulted in a program that failed to render a map, instead rendering only a legend. The potential for single prompts to have such dramatic impacts on the state of programs was disorienting and disconcerting to participants. As \p{8} reflected in the post-interview:

\blockquote{I became very frustrated by the fact that I could make all this good progress, and then I would put in a bad prompt or it would read it wrong, and then all of a sudden I've lost tether of that good progress and I can't really build on it \ldots{} I felt like I had no control of what was going on.}

Even when large-scale edits effected by \ghc{} did not lead to erring programs, they could have other negative impacts that made progress difficult. For example, \p{7} had successfully completed most of their \textsc{Reproduction} task with \ghc{}, developing a proportional symbol map that sized point locations of power plants in the United States according to their \code{total\_capacity} property. To wrap up, they issued a prompt to apply a categorical color scheme to the layer. While the resulting generated code correctly applied this change, it simultaneously \emph{removed} the proportional sizing of symbols (\p{7}: ``Well, we took one step forward and one step back here, Copilot.''). This potential to make progress on some fronts but regress on others again contributed to participants' sense of precarity when prompting; for experts, it was also a point where they abandoned \ghc{} in favor of direct text editing (\p{1}, \p{2}, \p{7}, \p{12}).

\subsubsection{Incremental Editing via Direct Manipulation in the \texorpdfstring{\hy{}}{DM+NL} and \texorpdfstring{\dm{}}{DM} Conditions Offered Continuity that Seemed to Affect Participant Understanding of Programs} \label{sec:incremental-editing}

Direct manipulation interfaces support editing via ``rapid, reversible, incremental actions'' \cite{shneiderman_1983} that allow users to ``see immediately if their actions are furthering their goals, and if not \ldots{} [to] change the direction of their activity'' \cite{hutchins_et_al_1985}. In the context of programming, we observed that this quality of direct manipulation also seemed to help participants reason about how edits they made to the program \emph{output} corresponded to transformations on the program itself. For example, while making changes to the \textsf{attribute} (\code{votes\_total} $\rightarrow$ \code{pct\_dem\_lead}) and \textsf{classification-method} (\textsf{Manual} $\rightarrow$ \textsf{Equal Interval}) parameters of a choropleth color scale in the GUI, \p{10} carefully observed how these edits altered the numeric values in the associated \code{fill-color} expression in the generated JavaScript code. Many participants employed a similar strategy of observing program updates when adding a new data layer to the map (\p{5}, \p{6}, \p{16}), switching map projections (\p{13}), or transitioning layer types (\p{5}) in the GUI. As \p{8} explained, the strategy was a useful tactic for bootstrapping program understanding:

\blockquote{I really like the fact that I can see the code live updating as I'm putting in new values. That feedback is so good for someone like me who eventually would like to know how to code using MapLibre. Having that instant feedback is really, really helpful as a learner.}

\noindent Importantly, the \emph{incremental} nature of direct manipulation edits appeared to be critical in this process. Because direct manipulation edits corresponded to conceptually discrete changes on the output that participants initiated one at a time, many could successfully identify the set of locations in the generated JavaScript program that had changed in response to their interaction.

\paragraph{\texorpdfstring{\textbf{Comparison}}{Comparison}: In the \texorpdfstring{$\nl{}$}{NL} condition, participants attempted to \texorpdfstring{``incrementalize''}{"incrementalize"} natural language edits}

Interestingly, several participants tried to emulate the incremental nature of direct manipulation editing in their interactions with code generated by \ghc{}. The most common behavior involved a form of manual program slicing, where participants copied only relevant ``slices'' of the generated code (e.g., a function definition and its call sites) from \ghc{} to their working environment in StackBlitz (\p{10}, \p{14}, \p{1}, \p{12}, \p{15}). Because this strategy required being able to (i)~identify particular slices in a large program based on a specific sub-goal and (ii)~trace slices to extract required dependencies, it came with risks; both \p{10} and \p{1} introduced errors through manual slicing that were not present in the generated code. But participants also noted that this slicing strategy came with the benefit of forcing incremental verification of small parts of the program, which in turn helped them understand program structure and localize sources of unexpected behavior (\p{10}, \p{1}, \p{12}, \p{15}).
Experts also explained that this technique was an attempt to recover patterns of how they already write code using external resources like StackOverflow or examples from online documentation.
\p{15}: 

\blockquote{It [\ghc{}] is throwing a lot at me, and I think that when I'm coding I like to kind of go line-by-line and like—if I'm pulling code over from another source I usually will pull things over in small little chunks to make sure I know which part is breaking it. Like, kind of break it as I go and fix it as I go. So it's challenging to pull over an entire script and then try to diagnose the problem.}

\noindent Some participants also tried to ``incrementalize'' outputs from \ghc{} by constraining the LLM at the prompt level. For example, \p{12} explicitly requested only a snippet of code as a response from \ghc{} in lieu of full program regeneration: \Prompt{Give me JUST the color scale stuff.}

\subsection{What Caused Participants to Switch between Direct Manipulation and Natural Language When Both Were Available?} \label{sec:switching-between-dm-and-nl}

\begin{figure*}
    \centering
    \includegraphics[width=0.95\linewidth]{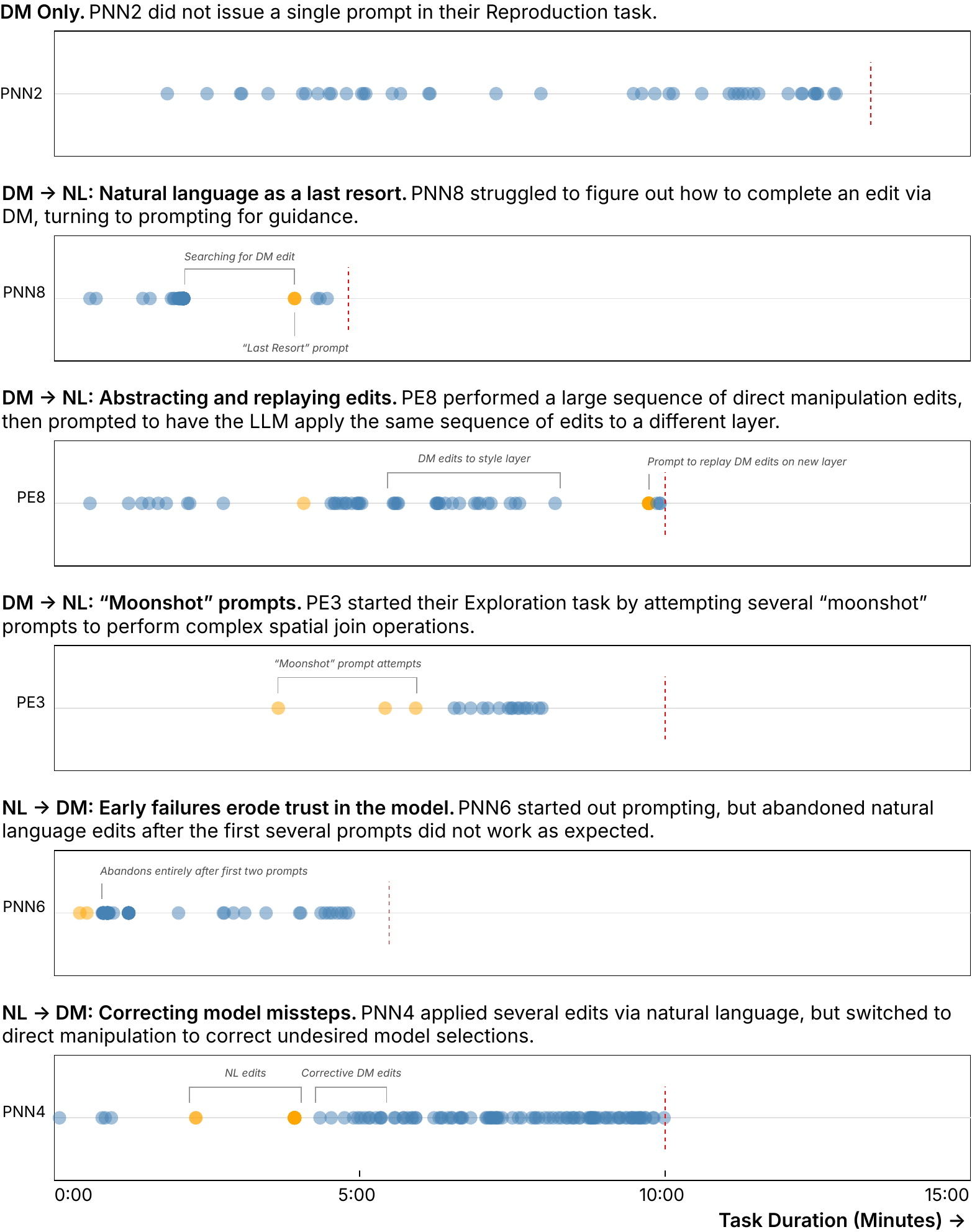}
    \caption{Examples of participants switching between direct manipulation \textcolor{SteelBlue!75}{\huge\textbullet} and natural language \textcolor{Orange!75}{\huge\textbullet} edits in \ckhy{}. Dashed red lines indicate when the participant successfully completed the task.}
    \Description{A series of strip plots showing six participants' direct manipulation and natural language edits over the course of a study task.}
    \vspace{-1em}
    \label{fig:interaction-plots}
\end{figure*}

The simultaneous availability of direct manipulation and natural language as program editing paradigms in \ckhy{} allowed us to examine when, how, and why programmers switched between the two. Based on both direct observation and telemetry data (Figure \ref{fig:interaction-plots}) of direct manipulation and natural language edits, we articulate several patterns of use.

\paragraph{DM Only: Some participants never switched to \texorpdfstring{$\nl{}$}{NL}}

Several participants elected to use direct manipulation only and did not issue a single prompt to the LLM in the course of completing the \textsc{Reproduction} (\p{6}, \p{3}, \p{12}, \p{15}) or \textsc{Exploration} (\p{5}, \p{11}, \p{14}, \p{8}, \p{12}) task. The most common explanation for avoiding the LLM from participants was that they already knew how to achieve their particular aim with direct manipulation, so turning to the model felt like an unnecessary detour. \p{12} explained:

\blockquote{I definitely did forget that [the LLM] was a feature, but also, looking back, I don't know what I really would've needed it for because I was directly manipulating the map and I didn't need help to figure that out \ldots{} I thought I could just accomplish everything without having to declare bankruptcy and be like, `You [LLM] tell me how to do this, I don't really know.'}

\paragraph{DM\texorpdfstring{$\rightarrow$}{→}NL: Some participants used \texorpdfstring{$\nl{}$}{NL} as a last resort}

A small handful of participants struggled to discern how to perform a particular program edit via direct manipulation in \ckhy{} and, as a last resort, turned to the LLM for help (\p{10}, \p{14}, \p{16}). For example, \p{16} wanted to map the \code{fill} channel of their active layer to a categorical color encoding using the (categorical) \code{primary\_source} property on their dataset. However, at the time of edit, the layer was set to use a quantitative color encoding; in this state, the \ckhy{} GUI preemptively filters out non-quantitative variables. Puzzled that \code{primary\_source} did not appear as an option in the GUI, \p{16} turned to \ckhy{}'s LLM: \Prompt{How can I show the power source of each power plant on the map?} GPT-5 correctly sequenced three \diff{}s in response to this prompt to modify the program and output to use the categorical color encoding.
In response, \p{16} exclaimed: ``Aha! Wow! It just did it for me, it didn't tell me how \ldots{} Cool, yeah, I couldn't figure that out.''

Still, many participants had quite high tolerance for struggling with direct manipulation in \ckhy{}, opting to stick with it even when interface navigation was a core challenge. For example, \p{6}, who took the second longest to complete the \textsc{Reproduction} task using \ckhy{} and had the most difficulty making direct manipulation edits, did not issue a single prompt in the course of the task. When asked why they did not turn to the LLM for assistance, they explained: ``I wasn't sure how to prompt it. I wasn't sure if it [the prompt] had to be in technical terms or if you could just say whatever and it still worked.''

\paragraph{DM\texorpdfstring{$\rightarrow$}{→}NL: Participants used \texorpdfstring{\nl{}}{NL} to automate or replay repetitive tasks}

After participants developed an understanding of how to complete larger-scale program edits via repetitive direct manipulation interactions, some turned to natural language to automate similar edits (\p{10}, \p{11}, \p{13}, \p{4}, \p{7}) or replay edits on new data (\p{9}, \p{10}, \p{4}, \p{8}, \p{15}). An example of the former came up when participants had to set specific thresholds for quantitative breaks in the color scale of a choropleth map, requiring six to nine direct manipulation interactions with multiple \code{<input>} elements. This kind of tedious, repetitive work was emblematic of tasks participants wanted to offload to the LLM. Natural language also allowed participants to leverage higher-level descriptions of the repetitive task rather than specifying precise literal values. For example, \p{7} directed the LLM to set breaks at \Prompt{integer values in 10 percentage-point increments moving away from zero} rather than specifying each threshold individually.

The latter use case---replaying edits on new data---involved participants trying to use the LLM almost like a parameterized record-and-replay or programming by demonstration system \cite{chasins_et_al_2018, barman_et_al_2016, chasins_et_al_2015}. In these instances, participants often spent significant time adjusting the symbology of one layer via direct manipulation, and then prompted the LLM to copy the symbology to a second layer while accounting for differences in the underlying datasets. The goal, again, was to avoid repeating many direct manipulation edits and instead delegate them to the model. \p{10} explained: 

\blockquote{Because the clickable interface is so easy to navigate, I only really felt compelled to [issue a prompt] when there was something tedious in front of me \ldots{} Like, I wanted to make the 2020 and 2024 [election data layers'] color bars the same. Like, ugh, that's gonna be a pain. That's gonna be a perfect, like, ask an LLM to change it.}

\noindent Prior work has speculated that this delegatory pattern of use could be a primary benefit to systems blending direct manipulation and natural language \cite{frohlich_1993}. While our empirical evidence confirms this, participants also expressed an extremely low tolerance for model errors in this context, explaining that almost any amount of error would obviate the automation benefit (\p{5}, \p{10}, \p{4}, \p{15}). 

\paragraph{DM\texorpdfstring{$\rightarrow$}{→}NL: Participants used \texorpdfstring{\nl{}}{NL} for \texorpdfstring{``moonshot''}{"moonshot"} prompts}

Particularly on unbounded \textsc{Exploration} tasks, several participants turned to \ckhy{}'s LLM to attempt ``moonshot'' prompts---computation that they presumed was not possible through direct manipulation but thought could be plausibly handled by an LLM. These tended to be more complex geospatial analysis operations that can be orchestrated (with significant expertise) in full-fledged desktop GISs, spatial databases, or Python libraries, including spatial or tabular joins across layers (\p{3}), computation of aggregate values or indices across layers (\p{16}, \p{3}, \p{4}, \p{7}), or bivariate symbologies (\p{7}). Some participants indicated that they did not necessarily expect these ``moonshot'' prompts to work, but used them as a proactive way to ``temperature check'' (\p{4}) or calibrate the LLM's capabilities. 

\paragraph{NL\texorpdfstring{$\rightarrow$}{→}DM: Early failures eroded participants' trust in the \texorpdfstring{\nl{}}{NL} tooling}

Many participants abandoned using \ckhy{}'s LLM entirely after encountering failures (\p{5}, \p{10}, \p{13}, \p{14}, \p{1}, \p{2}, \p{3}, \p{4}, \p{7}, \p{8}), whether that came in the form of the reserved \UnknownName{} \diff{} (signaling that GPT-5 could not construct a \diff{} sequence based on the prompt) or a timeout on response from the model. Some participants' tolerance for failure was extremely limited---even one failure was enough to incentivize abandoning. As \p{13} explained, success or failure of the first prompt had an outsized impact on participants' perception of model capability and utility: ``The first time you try to use it is such an important evaluation point, y'know? `Cuz I was like, `I'm gonna try this' and then it's like, `Error.' and I was like, `Ok, not using that.'{''} This is consistent with findings in prior work that just one or two failures are sufficient for users to presume model incapability \cite{zamfirescu-pereira_et_al_2023}. While other participants were more persistent in model use and engaged in systematic testing, we still observed consistent drop-offs in prompting over the course of tasks. Our hypothesis is that, in many instances, direct manipulation provided a comparatively easy pathway to progress---often even easier than direct text editing---when the model did fail, encouraging even quicker abandonment.

\paragraph{NL\texorpdfstring{$\rightarrow$}{→}DM: Participants used \texorpdfstring{\dm{}}{DM} to correct or repair model missteps}

Due to the inherently ambiguous nature of natural language, it was common for the \ckhy{} LLM to partially succeed---that is, it applied some of the \diff{}s the participant expected, but either failed to apply others or applied undesired \diff{}s. In these instances, participants turned quickly to direct manipulation to correct the model's missteps (\p{5}, \p{10}, \p{16}, \p{7}, \p{15}). For example, \p{15} attempted to transfer the symbology of one layer to another using natural language, similar to \p{10} above (\Prompt{Make the 2020 swing states layer match the symbology of the 2024 swing states, adjusting for the fact that the pct\_dem\_lead is scaled differently.}). While the LLM managed to match the symbology of the two layers quite closely, \p{15} could tell by looking at the updated map that certain discrepancies remained; in this instance, \ckhy{}'s LLM had failed to apply three \diff{}s to adjust the \code{fill-opacity}, \code{method}, and \code{scheme-direction} of the target layer. They turned immediately to direct manipulation to apply these lingering edits that the LLM had missed.

\subsection{Experience with Direct Manipulation Editing Helped Mitigate Known Challenges with Natural Language Editing} \label{sec:experience-with-dm-affects-performance-with-nl}

Two of the most well-studied obstacles identified in the literature on human-LLM interactions in natural language programming contexts involve (i)~inferring model capabilities \cite{liu_et_al_2023, nguyen_et_al_2024, zamfirescu-pereira_et_al_2023} and (ii)~interpreting model outputs \cite{obrien_2025, zi_et_al_2025}. In this section, we focus on how participants' experiences with direct manipulation helped mitigate these challenges in both the \hy{} and \nl{} conditions.

\subsubsection{Prior Exposure to the \texorpdfstring{\hy{}}{DM+NL} and \texorpdfstring{\dm{}}{DM} Conditions Improved Success of Prompting in the \texorpdfstring{\nl{}}{NL} Condition}

We observed that participants who used either \ck{} interface first in their study session often decomposed their natural language prompts in \ghc{} along \ck{}'s abstraction boundaries, referencing concepts like layer types, encoding channels, and classification methods (\p{6}, \p{9}, \p{10}, \p{11}, \p{13}, \p{17}, \p{2}, \p{4}, \p{7}, \p{8}). For example, \p{2}, who interacted first with \ckhy{}, issued the following initial prompt to \ghc{}:

\begin{quote}\small
    \Prompt{Make a choropleth map using the following US unemployment data. Use the rate property to determine the darkness of the polygon color. Use red. Darker colors should correspond with higher numbers. Bin the colors along quintiles. <Data URL>}
\end{quote}

\noindent The structure of this prompt maps almost directly to the order in which \p{2} interacted with direct manipulation controls in the \ckhy{} interface in the prior task, which they used to specify a layer type (\Prompt{choropleth map}), a GeoJSON property to map to the \code{fill} channel (\Prompt{Use the rate property to determine the darkness of the polygon color.}), a color scheme and direction for the \code{fill} channel (\Prompt{Use red. Darker colors should correspond with higher numbers.}), and a classification method for binning continuous numeric values into discrete bins (\Prompt{Bin the colors along quintiles.}).

Our hypothesis based on this behavior is that \ckhy{}'s direct manipulation GUI provided participants with effective scaffolding on \emph{how} to decompose a target map at a particular abstraction level, which they learned quickly enough to apply in a natural language setting. \p{10} confirmed this, describing \ckhy{} as a ``killer teaching tool'' that helps direct users when they have only ``half a picture of what they're trying to get out of their map.'' Beyond map decomposition, \ck{}'s GUI also gave participants access to technical cartographic and geostatistical terminology that they later used in prompts, particularly to describe the class of visualization (e.g., choropleth, dot density, proportional symbol) they sought from the LLM. Providing both a decomposition and technical language appeared to be especially valuable for near-novices; of the four near-novices who completed the \textsc{Reproduction} task with \ghc{}, three did so after seeing \ckdm{} or \ckhy{}. The fourth, despite being a near-novice in our study, is a professional cartographer who presumably already has access to both map decomposition skills and technical terminology.

\subsubsection{Participants Used the GUI in the \texorpdfstring{\hy{}}{DM+NL} Condition to Make Guesses about \texorpdfstring{``Allowable''}{"allowable"} Prompts, Improving Success} \label{sec:guessing-allowable-prompts}

Participants expressed uncertainty when prompting the LLM in \ckhy{}, citing confusion about ``what the model can see'' and ``[\textit{hovering over program, map, and GUI components}] what interface [is] the recipient of my prompts'' (\p{7}). Lacking knowledge of system internals, several participants looked to direct manipulation features in the \ckhy{} GUI to infer the kinds of program edits they could ask the LLM to make (\p{5}, \p{9}, \p{1}, \p{4}, \p{7}). Typically, this took the form of directing the LLM to perform edits that mapped very directly to specific GUI components. For example, one of \p{9}'s prompts---\Prompt{can you size the electricity sources points with graduated symbols, using the `total\_capacity' attr as the size}---maps directly to clicking the \code{electricity\_sources\_\_1} layer on the map, adjusting its \textsf{Layer Type} via a \code{<select>} element, and adjusting the attribute on its \textsf{size} channel also via a \code{<select>} element. When asked about how they reasoned about prompting the LLM in \ckhy{}, \p{9} explained:

\blockquote{Knowing what the tool was capable of before I started prompting was important. I wasn't doing joins and different spatial analysis things because I didn't see those tools in [the GUI]. I was constrained by what I knew I could do without the chatbot.}

\noindent In several cases, we observed participants keeping the layer editing controls visible \emph{while} writing their prompts, possibly as a reference aid (\p{10}, \p{15}). We hypothesize that this property of direct manipulation interfaces---that they ``announce'' what is possible through the interface itself, thereby providing a ``language'' of interaction---can help bridge the ambiguity gap between users and models.

\subsubsection{Participants Used the GUI in the \texorpdfstring{\hy{}}{DM+NL} Condition to Interpret Model-Triggered Edits} \label{sec:gui-to-interpret-model-triggered-edits}

When participants did turn to the LLM to perform certain program edits, \ckhy{}'s GUI was essential in helping them determine what the model did (\p{9}, \p{10}, \p{16}, \p{7}, \p{15}). For example, after issuing the prompt \Prompt{Can you bin the data in `2024 election results' by increments of 0.25?} and seeing their map update, \p{9} turned to the GUI to understand what edits the LLM had applied. Nearly immediately, they could tell the edits were only partially successful: ``Oh! Ah \ldots{} well, ok, something changed \ldots{} Oh, ok, so it switched it [the classification method] to \textsf{Equal Interval}. Oh, cuz maybe, I did say the word `increment.'{''} Understanding model edits through the GUI was effective for participants for two reasons. First, changes in the GUI were often easier for participants to quickly check for correctness compared to examining the generated JavaScript program. Interestingly, while participants also examined changes to the map to understand model edits (\p{5}, \p{2}, \p{7}, \p{15}), they often were not able to understand the scope of edits from the map alone. Second, as the above anecdote illustrates, participants used their understanding of model-triggered edits from the GUI to develop hypotheses about how their natural language prompts were being interpreted by the system. \p{7} described this process, sometimes called ``abstraction matching'' in the literature \cite{liu_et_al_2023, sarkar_et_al_2022}, as ``learn[ing] their [the LLM's] rules through some process of discovery,'' and noted that the GUI was particularly useful for this process.

\subsection{How Did Participants Assess Correctness?} \label{sec:how-did-participants-assess-correctness}

Researchers and industry practitioners have targeted programming domains producing visual outputs---such as UI design \cite{petridis_et_al_2023, petridis_et_al_2024}, web development \cite{vercel_v0}, and data visualization \cite{setlur_et_al_2016}---as good fits for natural language programming, presuming that outputs alone are sufficient for non-experts to assess program correctness. While we did observe that output examination was the primary way participants proxied for correctness in all conditions, we also found that this tactic was \emph{not} always sufficient, a finding in line with a recent study of how domain experts program with LLMs \cite{obrien_2025}. In this section, we discuss challenges participants experienced gauging program correctness.

\subsubsection{Soundness Guarantees in the \texorpdfstring{\hy{}}{DM+NL} and \texorpdfstring{\dm{}}{DM} Conditions Allowed Participants to Focus on the Output} \label{sec:soundness-guarantees}

The \ckhy{} and \ckdm{} \diff{}s are designed so that programmers cannot reach invalid programs in these systems and, consequently, cannot reach a state where the program output (i.e., the map) is not visible. For this reason, participants never entered an obvious debugging phase using these systems, nor did they express concerns about the correctness of generated code. Instead, ``debugging'' often manifested as choosing one program among a (theoretically infinite) space of valid programs constrained by the set of available \diff{}s. For this purpose, participants could (and primarily did) rely solely on examining the program output and the GUI state. 

Occasionally, changes to the program output triggered by a direct manipulation interaction contradicted participant expectations, triggering closer examination of the program (\p{5}, \p{3}, \p{4}). For example, after applying a proportional symbol transform to a dataset of US counties with information on broadband access rates, \p{5} noticed that the resulting symbols all appeared to be the same size despite expecting them to be variable. Confused, they turned their attention to the program. Through careful examination, \p{5} noticed that \ckhy{} was using a linear scale to size symbols, with endpoints at the minimum and maximum values of the \code{broadband\_access} variable. Using this information, they inferred that the distribution of \code{broadband\_access} was likely concentrated at the midpoint of the scale, such that many symbols would appear equally sized. Participants relayed that the relatively small size of generated programs in the \hy{} and \dm{} conditions aided in achieving this degree of code comprehension (\p{5}, \p{4}, \p{8}). 

\paragraph{\texorpdfstring{\textbf{Comparison}}{Comparison}: In the \texorpdfstring{\nl{}}{NL} condition, the model could (and did) generate programs that produced no visible output, leaving near-novices stranded}

Over half of participants reached some point where code generated by \ghc{} produced no output at all, commonly due to runtime errors such as incorrectly invoking a library API, referencing a non-existent column name on the dataset, or erroneously transforming dataset geometry. In these cases, participants knew their programs were incorrect but, absent a concrete output to refer to in subsequent prompts, had little way of communicating what was wrong to the model. This was a particularly large obstacle for near-novices, who typically resorted to notifying \ghc{} that its code did not produce visible output (e.g., \p{17}: \Prompt{The code doesn't work for me, it didn't show the preview appropriately.}), followed by restating the previous prompt in slightly different language (\p{6}, \p{11}, \p{17}, \p{15}) or asking the model to self-debug (\p{6}, \p{13}). These follow-ups still tended to fail, as participants could not identify underlying errors (such as incorrect library API usage) that the model was unlikely to correct in the next round of generation. Even when \ghc{} included a natural language summary of edits it made as well as step-by-step debugging instructions for the participant to take, we observed that participants largely ignored this information.

\paragraph{\texorpdfstring{\textbf{Comparison}}{Comparison}: In the \texorpdfstring{\nl{}}{NL} condition, the volume of generated code inhibited participants' ability to assess correctness}

Over half of participants also complained about the volume of generated code returned on every prompt in the \nl{} condition, explaining that it made discerning whether the model had implemented the desired functionality more difficult. Often, code verbosity was related to functionality the participant had not explicitly requested, such as legends, tooltips, unnecessary dataset parsing or transformation, or reimplementations of common geospatial operations (e.g., bounding box computation). Experts noted that this ``fluff \ldots{} made it harder to debug'' (\p{2}) and that \ghc{} ``seemed to way overcomplicate things'' (\p{1}). For near-novices, the volume of generated code was both intimidating and increased skepticism of model success. In response to receiving a 403-LOC HTML file after prompting \ghc{}, \p{5} said, ``I'm scared.'' While watching >600 LOC stream in from \ghc{}, \p{6} lamented: ``Not convinced by this.''

Beyond adding the overhead of unrequested functionality, large-scale code changes made pinpointing true errors more difficult. Because participants had not written the program themselves, they lacked an understanding of the program structure that would support error localization. \p{4}:

\blockquote{Some of [the difficulty of working with LLMs] is that it just vomits a whole ton of code on you at once. So it's like you're looking at an unfamiliar codebase \ldots{} It was more mentally taxing to try and figure out why Copilot couldn't do what I wanted it to do than it would have been to write the code myself.}

\noindent This finding is consistent with observations of experts in \cite{barke_et_al_2023, vaithilingam_et_al_2022}. For near-novices, we found that encountering any error largely halted progress because participants (i)~struggled to identify and meaningfully articulate the error to the model and, thus, (ii)~were dependent on the model to successfully self-debug. Prompts like \Prompt{It's not displaying. Please fix that} (\p{13}) or \Prompt{the power plants are not showing up on the map} (\p{6}) became common at this point.

\subsubsection{Participants Had Concerns About the Scope and Completeness of Model-Triggered Edits in the \texorpdfstring{\hy{}}{DM+NL} Condition} \label{sec:scope-completeness-llm-edits}

We observed that the addition of natural language editing in the \hy{} condition---by virtue of producing non-deterministic \diff{} sequences---added to the challenge of assessing the scope and completeness of model-triggered edits relative to expectations based on the prompt (\p{5}, \p{10}, \p{4}, \p{7}, \p{8}). As \p{7} remarked, issuing a natural language edit came with the burden ``of hav[ing] to go back and confirm that [the LLM] did what you expected it to do, even if you have the visual indicator that maybe it got somewhat close.'' \p{15} described this overhead as introducing a ``cost-benefit analysis'' to natural language editing, where they had to choose between the benefit of automating larger-scale edits and the cost of having to verify those edits. We also observed that, despite the multiple ``lenses'' participants had on natural language edits (in the form of changes to the program output, the program, and the GUI state), it was still easy to miss subtle incongruities between their prompts and updated outputs. For example, \p{4} issued a prompt to \Prompt{Style the 2024 layer as a choropleth by the \code{pct\_dem\_lead} fields [sic]} but, given the plausible appearance of the updated map, failed to notice that GPT-5 had selected the \code{votes\_total} property for visualization instead. Conversely, participants in the \dm{} condition never expressed concerns about the scope or completeness of edits because all edits were explicitly programmer-triggered.

\section{Limitations and Threats to Validity} \label{sec:limitations-and-threats-to-validity}

\subsection{Emergence of Coding Agents}

A key challenge in conducting any study involving human-LLM programming interactions is the rapid pace at which new models and model-backed tools are being released. When we began running study sessions in August 2025, \ghc{} was a state-of-the-art natural language programming system, built on a leading frontier model for programming (GPT-5) and equipped with tool use capabilities similar to coding agents (e.g., inspecting web pages via Playwright's MCP server~\cite{github-playwright-mcp}, reading public GitHub repositories). However, at the time of writing, new frontier models have since come out, and coding agents like \claude{}~\cite{claude_code} and \textsf{Codex}~\cite{codex} are popular. Expecting new model and tool releases, we aimed to focus our analysis on findings connected to the programming paradigms themselves (\hy{}, \dm{}, \nl{}) rather than the capabilities of any particular model or tool. For example, many participants remarked on GPT-5's lengthy inference time, but since this is a characteristic of a specific model, we do not discuss it in our analysis. Thus, we expect our findings will remain relevant even as models evolve.

Still, it is natural to ask: Can coding agents complete the tasks used in this study, given the researcher-written task descriptions? To answer this question, we conducted an experiment similar to the one we used to assess \ghc{}'s capabilities before our study (\S\ref{sec:experiment-design}, \textit{Task A}): using the full \textsc{Reproduction} task description (including screenshots of the target maps, URLs to data sources, and URLs to the original news articles) as input to a coding agent and seeing if it could complete the task according to our rubric. We ran this experiment in July 2026, using \claude{} running Opus 4.8 with High Effort on an M4 MacBook Pro with \qty{48}{\giga\byte} RAM.
We initially attempted to use Fable 5 as the underlying model, but hit cases where its safeguards blocked model responses.
We tracked the agent's token usage via \code{ccusage} \cite{ccusage}, as well as the wall-clock time.

\claude{} completed two of the three \textsc{Reproduction} tasks with no additional input beyond the researcher-written task description, but only completed the third task with guidance from the first author. In this third case, the agent failed to configure Google Chrome's Headless mode with a graphics backend that supported WebGL rendering with adequate performance for displaying the spatial datasets used in the task. The result was that screenshots the agent took were insufficient to assess the impacts of successive edits, leading to several issues. For example, the agent incorrectly styled one of the map layers and applied an incorrect spatial data transformation to features in a second layer. Using knowledge about the particular datasets and task, the first author was able to identify these issues and guide the agent through a fix in two separate prompts.

Completion times for the tasks were 6m36s, 11m30s, and 20m36s, and the agent used 612,510; 1,398,521; and 2,707,582 tokens, respectively. By wall-clock time, \claude{} took longer than study participants using \ckhy{} and \ckdm{}, but less time than participants using \ghc{}. However, note that, because we pasted the already-written task descriptions and screenshots directly into \claude{}, completion times do not include the time spent authoring the initial prompt; this time \emph{is} included in participants' timing data. Still, we speculate that, holding the study tasks and descriptions constant, using a coding agent for the \nl{} condition would bring task completion rates and times closer to those we observed in the \hy{} and \dm{} conditions. Additionally, certain classes of errors we observed with GitHub Copilot (e.g., those catchable by type systems or linters) would likely disappear almost entirely, though providing stronger correctness guarantees on the behavior of LLM-generated code remains an open area of research \cite{yang_et_al_2026, cai_et_al_2025}.

Because our study was designed to answer questions about programmers' processes, we selected tasks that were achievable within the allotted time with each tool while still requiring human participation.
For a variant of our study using different tools, the set of tasks that meet these criteria would be different (e.g., with a \claude{} condition, the third \textsc{Reproduction} task may be valid, while the other two would not be).
If future studies select tasks using the same criteria---achievable with all tools, but not fully automatable by any tool---we speculate that the findings will mirror our own.
Additionally, given our focus on programmer processes rather than model performance, we suspect that better models alone will not fully address the characteristics of our participants' \emph{interactions} that led to difficulties in the \nl{} condition.
For example, while cases where an agent introduces regressions (\S\ref{sec:scaffolding-edits}) or fails to produce visible program output (\S\ref{sec:soundness-guarantees}) may decrease, these improvements may not aid programmers in interpreting the scope, bounds, or desirability of agent-triggered edits (especially without, as in the \hy{} condition, an alternative for scrutinizing and amending these edits, \S\ref{sec:gui-to-interpret-model-triggered-edits} and \S\ref{sec:scope-completeness-llm-edits}). Open problems like these signal further opportunities for research blending agents' code generation capabilities with direct manipulation interfaces. 

% Moving forward, we encourage the community to bear in mind that task selection should vary based on the particular research question \emph{and} tools being investigated. When the tool \emph{du jour} is changing rapidly, comparisons that hold tasks constant may no longer help to answer the original research question. Further, researchers may wish to experiment with alternative instruction formats to increase the external validity of their studies and the chances that their findings will generalize to settings where detailed researcher-written instructions are not available.

\subsection{Domain Specificity}

We developed \ckhy{} for a particular domain (geospatial visualization) and evaluated it with practitioners with expertise in that domain. While this study design lends our findings ecological validity in this context, it also limits our ability to generalize our results to other domains, even those with established direct manipulation tooling. More likely still, the patterns we report may \emph{only} appear in domains where mature direct manipulation tools are widely adopted, such as image processing, vector graphics, computer-aided design, or 3D modeling. Given that, to our knowledge, all existing direct manipulation programming systems (e.g., \cite{mayer_et_al_2018, hempel_chugh_2016, hempel_et_al_2019, ziegler_et_al_2025, zhang_et_al_2024, zhang_et_al_2023, fukahori_et_al_2014}) are specialized to a specific domain, this is a limitation we share with prior work.

\subsection{Participant Preferences}

Our focus on data journalists as a study population---specifically those with experience working with geospatial data---could have biased the patterns of use we observed across the three conditions. As an example, our participants' preference for direct manipulation editing in the \hy{} condition may be connected partly to their existing tool use patterns. Responses from our post-study survey showed that all 18 participants had prior experience working with GISs or vector graphics software for their cartographic work---tools whose primary interaction model is direct manipulation---but just over half of them (11/18, see Table \ref{tab:participants}) had used an LLM-based natural language programming system prior to our study. Thus, our findings may not extend to other programmer communities where the surrounding tooling ecosystem is significantly different.

Interestingly, we found no relationship between participants' prior experience working with LLMs and their performance on \textsc{Reproduction} tasks, which suggests that challenges participants faced in the \nl{} condition are unlikely to be explained entirely by unfamiliarity with natural language programming systems. Specifically, we fit a logistic regression predicting \textsc{Reproduction} task completion in the \nl{} condition from participants' frequency of LLM use for programming, and found that the model fit poorly ($p = 0.72$). Even with low evidence for this model, its effect size is small (pseudo-$R^2 = 0.08$, well below the 0.2--0.4 range considered ``excellent fit'' \cite{mcfadden_1979} for generalized linear models such as a logistic regression model).

\section{Related Work} \label{sec:related-work}

\subsection{\texorpdfstring{\hy{}}{DM+NL} Programming Systems and User Studies}

To our knowledge, there are no programming systems supporting multimodal direct manipulation and natural language editing prior to this work, and thus no user studies of such systems.

\subsection{User Studies of Direct Manipulation Programming}

Direct manipulation programming is a rich area of research at the nexus of programming languages and human-computer interaction, with researchers developing direct manipulation programming systems for SVG \cite{chugh_et_al_2016, hempel_chugh_2016, hempel_et_al_2019, zhang_et_al_2024}, HTML \cite{mayer_et_al_2018, zhang_et_al_2023}, game development \cite{fukahori_et_al_2014}, and geospatial visualization \cite{ziegler_et_al_2025}. However, despite the proliferation of system development in this space, we are not aware of \emph{any user study of direct manipulation programming}. Thus, this work represents a first step in filling this gap in the literature. Interestingly, our results offer the first evidence for some of the presumed benefits motivating prior work (e.g., that direct manipulation programming can effectively scaffold the programming process for near-novice programmers) \cite{hempel_et_al_2019}.

\subsection{User Studies of Natural Language Programming}

Research on natural language programming dates back to at least the 1960s \cite{barnett_ruhsam_1968}, and the recent arrival of code-generating LLMs has driven widespread interest in understanding how these technologies shape the programming process. Studies in this space are numerous, with researchers conducting in-lab experiments, observational studies, randomized controlled trials in the field, evaluations of field deployments, and more, with expert programmers \cite{vaithilingam_et_al_2022, barke_et_al_2023, becker_et_al_2025, murali_et_al_2024, feng_et_al_2024, ziegler_et_al_2024}, novice programmers \cite{nguyen_et_al_2024, lucchetti_et_al_2025, denny_et_al_2024, lau_guo_2023, kazemitabaar_et_al_2024}, domain experts \cite{obrien_2025, wang_et_al_2025}, or some combination \cite{zamfirescu-pereira_et_al_2025, yan_et_al_2024}. 
Our work builds on these studies by examining how users' natural language programming practices change when we add direct manipulation as an editing modality. Most excitingly, our findings indicate that direct manipulation can alleviate key challenges with natural language programming, including issues with understanding model capabilities \cite{nguyen_et_al_2024, zamfirescu-pereira_et_al_2023} and interpreting a model's program edits \cite{obrien_2025}.

\subsection{\texorpdfstring{\hy{}}{DM+NL} Systems Outside of Programming Contexts}

Prior work has blended direct manipulation and natural language in interfaces for visualization editing \cite{vaithilingam_et_al_2024, wang_et_al_2023}, styling web pages \cite{kim_et_al_2022}, vector graphics \cite{masson_et_al_2024}, and animation \cite{bourgault_et_al_2025}. For example, \textsc{DynaVis} \cite{vaithilingam_et_al_2024} leverages users' natural language prompts to generate ``dynamic widgets,'' new pieces of UI to support direct manipulation of visualization properties. Their approach uses an LLM for code completion on strict, pre-defined HTML and JavaScript templates, followed by a lightweight program analysis. \textsc{DirectGPT} \cite{masson_et_al_2024} uses a combination of engineered prompts alongside instrumentation that adds contextual information to text or SVG. Our work differs from these prior systems in two ways. First and most importantly, these tools are not \emph{programming systems} and do not produce programs as primary outputs for users. Second, the approaches these systems employ rely on an LLM to generate \emph{unstructured} output (e.g., completions for function bodies), which provides no guarantees on correctness or compatibility with the host system. In contrast, our approach uses an LLM to generate sequences of \emph{structured} program edits, which we can design to (i)~guarantee correctness and (ii)~obviate the need for additional program analysis.

\subsection{Constrained Decoding in Programming Contexts}

There has been substantial research exploring how constrained decoding can improve the syntactic and semantic correctness of code generated by LLMs \cite{nagy_et_al_2026, mundler_et_al_2025, agrawal_2023, poesia_et_al_2022, beurer-kellner_et_al_2024} and even support higher-level program authoring through natural language \cite{beurer-kellner_et_al_2023}. In general, these tools do not use constrained decoding specifically for generating program \emph{edits} but rather focus on full program synthesis with constraints. Notably, however, these approaches are complementary to our work because \diff{}s are an edit \emph{language}. For example, a system developer could use a framework like \textsc{ChopChop} \cite{nagy_et_al_2026} as the constrained decoder to generate \diff{}s with extensive semantic constraints; similarly, in latency-sensitive settings like ours, a constrained decoder like DOMINO \cite{beurer-kellner_et_al_2024} could speed up \diff{} generation. Thus, we believe that advances in constrained decoding for code generation have substantial implications for improving LLM-backed systems built on \patch{}-\recon{}.

\subsection{LLMs for Program Editing}

Prior work has explored using LLMs for other forms of program editing, including automated program repair \cite{zhang_cambronero_et_al_2024, xie_et_al_2025, rahman_et_al_2025}, automated refactoring \cite{pomian_et_al_2024}, large-scale code migration \cite{ziftci_et_al_2025}, structure editing and live programming \cite{blinn_et_al_2024}, and edit prediction \cite{gupta_et_al_2023}. In addition, many industrial tools (e.g., \claude{} \cite{claude_code}, \ghc{} \cite{github_copilot}, \textsf{Codex} \cite{codex}) focus on program editing as well as drafting. In general, these systems still produce unstructured token sequences in the host programming language as outputs, as opposed to our approach of generating structured edits in an edit language.

\subsection{Programming by Demonstration}
\label{sec:relwork-pbd}

Participants' use of natural language to replay repetitive tasks (\S\ref{sec:switching-between-dm-and-nl}) connects to a long line of work on programming by demonstration (PBD). In PBD, users provide a trace of actions, and the system generates a program to generalize and replay these actions in new contexts. \citet{Lau03}'s work on SMARTedit uses user traces of repetitive text edits to prune a large \textit{version space algebra} of possible generalizations.
Since then, researchers have introduced a variety of techniques to generalize users' actions, including for web automation~\cite{Leshed08,Lin09,chasins_et_al_2018,Chen23,Li24}, data transformation~\cite{Kandel11}, and, most relevantly, repetitive code edits~\cite{Miltner19,Brody20,Ni21,Zhang22}. Recently, researchers have developed neurosymbolic systems that incorporate machine learning techniques and models into PBD algorithms~\cite{Chen21,Pu23,Patton24}, including many that leverage LLMs for repetitive code edits~\cite{Wei23,gupta_et_al_2023,Liu24,Liu26,Chen26}. Our participants' behavior of providing a GUI demonstration and using natural language to request a generalization provides motivation for work that blends natural language input with PBD's symbolic guarantees.

\subsection{Program Comprehension}

Participants' use of the GUI to understand their underlying program---both the space of possible modifications (\S\ref{sec:guessing-allowable-prompts}) and the edits already made (\S\ref{sec:gui-to-interpret-model-triggered-edits})---relates to prior work on program comprehension. \citet{Brooks83}'s influential model describes program comprehension as ``based on the successive top-down refinement of hypotheses about other knowledge domains and their relationship to the executing program''~\cite{Brooks83}. Our participants used the GUI to start their exploration of these hypotheses in what Brooks would describe as the ``knowledge domain'' of cartography in order to learn about the ``programming domain'' of JavaScript in a top-down fashion.
In this way, the GUI components that served as gauges of the program were analogous to Brooks's \textit{beacons}, fragments of code that signal ``the occurrence of certain structures or operations,''~\cite{Brooks83} which follow-on work has identified as playing a key role in expert programmers' program comprehension~\cite{Wiedenbeck86,Koenemann91}.
% Further in line with \citet{Brooks83}, rather than starting with what \citet{Soloway84} define as a \textit{programming schema} (which ``describe a global strategy used in a program or algorithm and specify language independent actions''~\cite{vonMayrhauser95}), participants started with domain-specific (i.e., cartographic) questions about their program when seeking to comprehend it.
% Finally, in contrast to the predictions of bottom-up or integrated models of program comprehension~\cite{Pennington87a,vonMayrhauser93,vonMayrhauser97}, we did not observe participants engage in reasoning about the underlying program's control and data flow. We hypothesize this discrepancy may be due to the presence of the direct manipulation GUI, which may center the top-down domain-based program comprehension strategies over others.

\section{Discussion and Conclusion} \label{sec:discussion}

\subsection{Why Are Edit Languages with \texorpdfstring{\patch{}-\recon{}}{patch-recon} a Good Fit for \texorpdfstring{\hy{}}{DM+NL} Programming?}

In this work, we presented an approach to building multimodal direct manipulation and natural language programming systems based on edit languages and the \patch{}-\recon{} framework. This is just one of many possible architectures, but our findings point to two key areas where it excels.

\paragraph{Stability of Direct Manipulation Editing}

In \S\ref{sec:soundness-guarantees}, we discuss how the proof of patch-reconciliation correspondence over \ck{}'s \diff{}s prevented participants from reaching invalid programs in both the \dm{} and \hy{} conditions, which in turn minimized time spent debugging. But this soundness guarantee carried an additional, less obvious benefit: regardless of the edits dispatched by the LLM, participants never reached a state where direct manipulation editing stopped working.

Consider an alternative design for a \hy{} programming system where the GUI is constrained to known edits but the LLM may rewrite the program arbitrarily. In this setup, the LLM could introduce new control flow, data structures, or stateful logic with no clear mapping to GUI controls; it could also violate invariants that existing direct manipulation functionality relies on. Based on how participants used direct manipulation editing, we suspect that such ``progressive degradation'' could have negative consequences for programmers. As we discuss in \S\ref{sec:guessing-allowable-prompts} and \S\ref{sec:gui-to-interpret-model-triggered-edits}, participants in the \hy{} condition relied heavily on the GUI both to structure and interpret LLM-triggered edits; this functionality would be partly or wholly unavailable in this alternative formulation. Likewise, as we discuss in \S\ref{sec:switching-between-dm-and-nl} (\nl{}$\rightarrow$\dm{}: Participants used \dm{} to correct or repair model missteps.), participants often switched from natural language to direct manipulation to fix cases where the LLM partially succeeded but, owing to natural language's ambiguity, missed some of their intended edits. Here too, the alternative approach would all but eliminate this capability. 
%Ultimately, our design was driven by the principle that a \hy{} programming system should aim to keep both editing modalities available at all times, and we chose to limit the LLM's flexibility as a consequence.

\paragraph{Performance of Natural Language Editing}

The original motivation for \patch{}-\recon{} in \citet{ziegler_et_al_2025} was to improve the interactive performance of direct manipulation programming systems in domains where program outputs are expensive to compute. Geospatial visualization fits this description, as do many others: computer-aided design, 3D modeling, animation, image processing, and simulation. \patch{}-\recon{} makes direct manipulation programming tractable in these contexts by giving an incremental semantics to program edits dispatched by GUI interactions.

By structuring natural language edits to integrate directly with \patch{}-\recon{} via constrained decoding, we bring these same performance benefits to natural language editing with minimal overhead (just network latency and model inference time). We hypothesize that this partly explains behaviors described in \S\ref{sec:preferring-dm} and \S\ref{sec:switching-between-dm-and-nl} (\dm{}$\rightarrow$\nl{}: Participants used \nl{} to automate or replay repetitive tasks.): automating large-scale or repetitive edits was worthwhile not only because direct manipulation was tedious in these contexts, but because natural language edits had similar performance characteristics to direct manipulation edits (though, as we discuss in \S\ref{sec:scope-completeness-llm-edits}, introduced more uncertainty). Again, the alternative described above would enjoy none of these benefits; any natural language edit would require full program re-evaluation, making interactive performance dependent on the execution time of the program as a whole.

\subsection{Is Natural Language Really Natural?}

Our results suggest future work should more deeply explore some of the motivations of prior work on natural language interfaces \cite{wang_et_al_2023, setlur_et_al_2016, li_et_al_2019, mihalcea_et_al_2006}, which claim natural language requires ``less prior knowledge'' \cite{wang_et_al_2023}, ``should have a low learning barrier for end users'' \cite{li_et_al_2019}, and ``drastically increas[es] the accessibility of programming to non-expert users'' \cite{mihalcea_et_al_2006}. While these claims sound reasonable, our results suggest a murkier picture. Although natural language programming may remove the obstacle of learning the syntax and semantics of a programming language, it may also add an entirely new obstacle of learning the subset of natural language that produces good results (or, as \p{7} called it, ``learning to play by [the model's] rules''). Assessing the ``naturalness'' of natural language is far from the central research question of this work, but our findings do reinforce the growing body of literature calling its naturalness into question \cite{liu_et_al_2023, nguyen_et_al_2024, lucchetti_et_al_2025, feldman_anderson_2024}.

\subsection{Direct Manipulation and Natural Language Programming, Together at Last?}

Our results offer a first glimpse of the benefits of combining direct manipulation and natural language programming, but they also demonstrate that direct manipulation can be a powerful programming paradigm in its own right. This is perhaps not surprising given the broad adoption of direct manipulation interfaces in everyday computer use, and yet we did not find empirical evidence for this finding in the literature. Pioneering work on direct manipulation programming \cite{chugh_et_al_2016, hempel_chugh_2016, mayer_et_al_2018, hempel_et_al_2019, zhang_et_al_2023, zhang_et_al_2024, ziegler_et_al_2025} developed the technical foundations of the paradigm but primarily evaluated expressiveness; building on those foundations, our work goes a step further by examining how the paradigm supports programmers \emph{in practice}. Our findings hint that direct manipulation programming may be able to live up to its promise of simplifying programming relative to alternative paradigms, and, in specific contexts, may even be programmers' preferred modality.

Still, we aim to move away from framings that position direct manipulation and natural language as oppositional approaches (e.g., \cite{shneiderman_maes_97}). Rather, we see them as complementary program editing paradigms that can coexist in increasingly sophisticated programming systems. This paper offers a first look at what writing code with such systems looks like for programmers, and provides a possible blueprint for their design based on the idea of a shared language of structured edits.

\section*{Data-Availability Statement}

\ckhy{} is open source and freely available at \url{https://github.com/parkerziegler/cartokit}, and the production deployment of \ckhy{} is available at \url{https://alpha.cartokit.dev}. Researchers interested in gaining free access to natural language editing on the production deployment may contact the first author directly; they may also build \ckhy{} from source using their own OpenAI API key. In addition to the GitHub repository, we provide an archived snapshot of \ckhy{}'s source code in a separate Zenodo archive \cite{cartokit_dm_nl_source}. Beyond source code, this archive contains (1) telemetry and survey data collected in the course of our study, including all participant timing data, prompts, \diff{}s, survey responses, and NASA-TLX scores; and (2) implementations of the statistical analyses and graphs in \S\ref{sec:quantitative-results}. Pre-built Docker images to reproduce all statistical analyses and graphs are also available on Zenodo \cite{cartokit_dm_nl_ae}.

\begin{acks}
    First and foremost, we are deeply indebted to our anonymous study participants and collaborators in the data journalism and geospatial research communities, whose generosity, time, and expertise made this research possible. We would also like to thank our anonymous OOPSLA reviewers for their feedback and deep engagement with our work. Members of PLAIT Lab and the EPIC Data Lab at the University of California, Berkeley provided valuable guidance that shaped this work throughout its lifecycle. Finally, a special thanks is due to the first author's dog, sweet Lou, whose patient companionship on many long walks led to the best ideas of this paper. This work was supported by CITRIS and the Banatao Institute, the College of Computing, Data Science and Society at the University of California, and by the Academic Innovation Catalyst. This work was also supported in part by NSF grants FW-HTF 2129008 and CA-HDR 2033558, as well as by gifts from Google, G-Research, Adobe, and Microsoft. Chasins is a Chan Zuckerberg Biohub Investigator.
\end{acks}

\bibliographystyle{ACM-Reference-Format}
\bibliography{references}

\end{document}